\documentstyle[aps,prd,preprint,tighten,floats,epsf]{revtex}

\def \pom {{\hspace{ -0.1em}I\hspace{-0.2em}P}}

\def \mb {{\rm mb}}

\begin{document}
%

\preprint{VAND-TH-99-10}

\title{A Tale of Two Hard Pomerons} 

\author{Arjun Berera\thanks{E-mail:
ab@ph.ed.ac.uk}
 \thanks{Present address, Department of Physics and
Astronomy, University of Edinburgh, Edinburgh EH9 3JZ, United Kingdom}}

\address{
Department of Physics and Astronomy,
Vanderbilt University, Nashville, TN 37235, U.S.A.}

\date {December 29 , 1999}

\maketitle

\begin{abstract}
Two mechanisms are examined for hard double pomeron exchange dijet
production, the factorized model of Ingelman-Schlein and
the nonfactorized model of lossless jet production which
exhibits the Collins-Frankfurt-Strikman mechanism.
Comparison between these two mechanisms are made of the total cross
section, $E_T$-spectra, and mean rapidity spectra. For both mechanisms,
several specific models are examined with the cuts of
CDF, D\O\, and representative cuts of LHC.  Distinct qualitative
differences are predicted by the two mechanisms for
the CDF $y_+$-spectra and for the $E_T$-spectra for all
three experimental cuts.
The preliminary CDF and D\O\ experimental data
for this process are interpreted in terms of these two mechanisms.
The $y_+$-spectra of the CDF data is  
suggestive of domination by the
factorized Ingelman-Schlein mechanism, whereas
the D\O\ data shows no greater preference
for either mechanism.  An inconsistency is found
amongst all the theoretical models in attempting
to explain the ratio of the cross sections given by the data
from these two experiments. 


\end{abstract}

\medskip

PACS number(s): 13.85.-t, 12.38.Qk, 13.87.Ce

\medskip

hep-ph/9910405

\bigskip

In press Physical Review D, 2000

\bigskip


\section{Introduction}
\label{sect1}

In diffractive hard scattering, the incident hadron
in e-p collisions and one or both hadrons in ${\bar p}p$ collisions
participate in a hard interaction involving very large momentum transfer,
but nevertheless the respective hadrons emerge with small
transverse momenta and a loss of small fractions of their longitudinal
momenta. 
For such diffractive hard processes, first comes a
question of pure semantics, whether or not to say the diffractive proton
"exchanged a pomeron".   Only one Pomeron
has entitled historical rights to this name, 
and that is the Pomeron of soft Regge
physics \cite{pomeranchuk,Regge:1959mz}
(also sometimes called the soft Pomeron). 
Reference to pomeron in any other
case exploits this established trademark as a mnemonic for describing
some portion of the process in which a strong interaction scattering
occurred that involved the exchange of no quantum numbers except
angular momentum.   In our discussion of diffractive hard scattering,
we will use the lower case pomeron in reference to
a process in which one or both incoming hadrons diffracts into the final
state along with a hard process.  On the other hand, the upper case
Pomeron will be reserved for the vacuum exchange trajectory of soft
Regge physics \cite{pomeranchuk,Regge:1959mz}.

There is general belief that properties of the Pomeron reflect in the
pomeron of diffractive hard scattering, although it is a central
research question to identify the specifics.   Spacetime arguments
generically suggest that hard events are well localized in space and
time.  Thus it is expected that in a diffractive hard process, the
diffractive hadrons undergo effects similar to what they would
encounter in a high-energy elastic scattering.  As such, diffractive
hard physics is expected to involve long-time, long distance,
thus nonperturbative, physics.  Nevertheless, that hard processes can
occur intermittent to the diffractive scattering indicates that
diffractive hadronic physics, via the pomeron,
also possesses perturbative properties that can be explained through
perturbative QCD. 

A primary goal of diffractive hard scattering physics is to unify the
QCD picture of the pomeron with the phenomenological Regge physics
description (for a review of Regge phenomenology 
applied to diffractive physics 
please see \cite{Alberi:1981af,Goulianos:1983vk}).  
Hard double pomeron exchange (DPE) processes
are useful in addressing this question, since it turns out
the QCD and Regge physics description of these processes have
some distinct qualitative differences, which are best expressed
in the context of hard factorization.

Recall, for a hard scattering factorized process, the effect of the
two incoming particles act independently 
on the hard event \cite{cssrev,css}.
The basic Regge physics motivated model of hard diffractive 
processes is the Ingelman-Schlein
model\footnote{Their model was motivated by a prior
and seminal diffractive hard scattering experiment by the UA4 \cite{ua4}
and subsequently the ideas of their model were first studied
by a UA8 experiment \cite{Bonino:1988ae}.
Some other theoretical works at around the same time
as this model also had similar 
ideas \cite{frst,Berger:1987iu}} \cite{Ingelman:1985ns}, 
and this model assumes
hard factorization.   
In their model, diffractive
scattering is attributed to the exchange of a pomeron, which
operationally is defined as a colorless object with vacuum quantum
numbers.  Their model treats the pomeron like a real particle and so
considers, for example, that a diffractive electron-proton collision
is due to an electron-pomeron collision and that a diffractive 
hadron-hadron collision is due to a proton-pomeron collision for
single-sided diffraction and pomeron-pomeron collision for double
diffraction.

For diffractive deep inelastic scattering, basic ideas 
of hard factorization were outlined
and diffractive parton distribution functions were defined
in \cite{Berera:1994xh,Berera:1996fj}\footnote{Closely
related to diffractive parton distribution functions
are fracture functions \cite{vene}}. 
A proof of
factorization for diffractive DIS was given in \cite{Collins:1998sr}.
For hard diffraction in pure hadronic collisions,
Collins, Frankfurt and Strikman (CFS) \cite{Collins:1993cv} have demonstrated
a counterargument to hard factorization. 
The CFS mechanism is a leading
twist effect in which all the momentum lost by the diffractive hadrons
goes into the hard event.   An important feature about the CFS mechanism
is that it requires the color flow properties of QCD in an essential
way.  In general, the presence of color in QCD implies pomeron exchange in
simplest form is a two gluon exchange process \cite{lownus,gunsop}. 
Necessarily, the simplest
model of the pomeron must involve at least two partons
in order to be color singlet. 
The two gluon pomeron model has a key
property for any pure hadron initiated reaction, which is a realization
of the CFS mechanism.  Consider the hard DPE process
$h_1 h_2 \rightarrow h_1 h_2 + hard$,
where $h_1,h_2$ are the colliding hadrons.  
The two gluons exchanged by $h_1$
are not both obliged to enter the hard event.  Instead, one gluon 
may attach to $h_2$.  In this case, the two incoming hadrons no longer
act independently in inducing the hard event.  By definition of hard
factorization \cite{cssrev,css}, such a process is nonfactorizing.  

This mechanism was identified 
earlier by Frankfurt and Strikman \cite{fs1}.  They originally
referred to the nonfactorized pomeron of CFS as the coherent pomeron. 
Subsequently, the UA8 presented results \cite{Brandt:1992zu}
in which up to 30 percent of the
dijet events in single-sided diffraction could be 
associated with the coherent
pomeron, which they in turn named the superhard pomeron.  With the
hindsight of the UA8 experiment and the ideas of CFS, in
\cite{Berera:1994xh} the CFS mechanism was applied to a toy quantum field
theory model of diffractive dijet photoproduction, in which the pomeron
was represented by two gluon exchange.  This work in turn, in turn,
named the nonfactorizing, alias superhard, alias coherent
pomeron process as lossless diffractive hard scattering to
emphasize the efficient transfer of the 
pomeron momentum to the hard process.

The CFS mechanism has been developed for hard DPE in ${\bar p} p$
collisions\footnote{The first
nonfactorizing DPE two gluon model was developed before CFS for Higgs
\cite{bl} and heavy quark \cite{Bialas:1992yc} production.  
Although the nonfactorizing
mechanism is the same as that of CFS \cite{Collins:1993cv} and 
\cite{Berera:1994xh,pumplin,Berera:1996vi}, these
earlier papers did not recognize the full consequences of
nonfactorization to the extent done by CFS.}
for quark jets in \cite{pumplin} and gluon jets
in \cite{Berera:1996vi}. 
The gluon jet process was shown in \cite{Berera:1996vi} to dominate
the quark jet process by several orders of magnitude.

The purpose of this paper is to examine for the DPE dijet process,
general differences between the factorized pomeron model of
Ingelman and Schlein, F(IS)DPE \cite{Ingelman:1985ns}, 
and the nonfactorized pomeron model
of lossless jet production of Berera and Collins \cite{Berera:1996vi},
N(L)DPE.  Our notation specifies in the context of hard factorization
whether the process is factorizable, F, or nonfactorizable, N,
and in parenthesis gives the particular type of process.
The latter specification is necessary
since there are several different
types of factorizable and nonfactorizable processes. 
Detailed discussions about this point
are in 
\cite{Collins:1993cv,Berera:1994xh,Berera:1996fj,Berera:1996vi,Collins:1998sr,hks,hebecker}.
As one example, factorized processes first have a basic
distinction between simple hard factorization and the more specific
Regge factorization \cite{Berera:1996fj}.
In particular, the factorized Ingelman-Schlein DPE model also is
Regge factorized\footnote{Hereafter our usage of factorization
without further specification always means in the context of hard
factorization.}.  

For nonfactorization, one example outside of the CFS mechanism is the "flux
renormalization" prescription of Goulianos \cite{kg1}, which
arises due to a breakdown of the triple-Regge theory for soft
diffractive excitation.  Also, nonfactorization is found in pre-QCD
analysis of diffractive processes \cite{preqcddiff}.
An empirical analysis by Alvero, Collins and 
Whitmore \cite{Alvero:1998dw}
of the preliminary CDF double
diffractive dijet data \cite{cdfdd1,cdfdd2} 
indicates that hard factorization is
violated in this process. In fact, their analysis suggests 
for parton
distribution functions that are most consistent over all diffractive
processes, the experimental DPE dijet cross section 
is much less than expected by
factorization.  On the other hand, the nonfactorizing CFS mechanism
should enhance the cross section.  Nevertheless, the
analysis in \cite{Alvero:1998dw} does not rule-out 
experimental realization of the
CFS mechanism, since general understanding from Regge models suggests
that there is a large source of suppression 
which will emerge from effects generically termed absorptive
corrections.  These effect are due to exchanges of pomerons and
gluons between particles in the basic model that possess
very different rapidities, thus in particular between the two incoming
hadrons.  As such, these effects also are nonfactorizing. Actual 
computation of absorptive corrections is nontrivial since they are 
nonperturbative.  Some work has been done to estimate their effects
\cite{regmod,kmrabsorb}. A general conclusion of these works is that
absorptive correction effects are independent of the
hard kinematics and weakly s-dependent.  As such, these effects
should be very easy to distinguish from the N(L)DPE
process. Also, these effects only should shift, in
particular decrease, the values of the cross sections from those
computed in our basic models and the effect should be the same
for either the F(IS)DPE or N(L)DPE processes.  In this paper,
we are interested in examining qualitative differences
between the F(IS)DPE and N(L)DPE processes, which are
minimally model dependent.  For this we will examine
the $E_T$ and mean rapidity ($y_+$) spectra for both  processes
and for the
cuts of CDF, D\O\ and
representative cuts for LHC.
We also will present total cross sections for all
the models and all the experimental cuts.  Thus the interested
reader can test any suppression factor from
any absorptive correction model that they wish.

The reason that we do not give a demonstrative example
of the overall absorptive correction suppression 
factor is that, as will be seen
in the sequal, for all the models that are examined, we find 
disagreement in the ratios of the cross sections from
those found in the available experimental data.
This discrepancy minimally is of order $\sim 5$.
Present understanding about absorptive corrections can not
explain this discrepancy, since their effect only
is to shift the cross sections by the same overall correction factor
which drops out in the ratios. 
This discrepancy may reflect upon 
a limitation of our partonic level calculations
or other controllable theoretical sources, or it
may be that since the experimental data is still preliminary,
it may yet be modified.   We will not attempt
to formulate any theoretical explanations for
this discrepancy found in this paper.  Our modest goal is
to examine the predictions of the basic models, which
up to now still have not adequately been done for these processes.
Of special interest is to identify  features that 
are minimally model dependent.  Furthermore, in 
light of the breakdown of
hard factorization suggested in \cite{Alvero:1998dw},
it is important to know whether any features of the basic models
are seem in the data.

The paper is organized as follows.  Sect. \ref{sect2}
reviews the kinematics of DPE dijet production and then models
are presented for the nonfactorized and factorized processes.
The inclusive dijet cross section also is reviewed in Sect. \ref{sect2}
and will be computed in later sections for comparison purposes.
In Sect. \ref{sect3} the CDF and D\O\ cuts for DPE dijet production
are reviewed and representatives cuts for LHC are presented.
In Sect. \ref{sect4} results of our calculations are presented for
DPE and inclusive dijet $E_T$ and mean rapidity spectra and total
cross sections.  Subsect. \ref{subsect4A} gives a general presentation
of the results, Subsect. \ref{subsect4B} examines the results in 
greater detail and Subsect. \ref{subsect4C} compares our
results with the preliminary data of CDF and D\O\ .  Finally
Sect. \ref{sect5} presents our conclusions.
Also in the last part of Sect. \ref{sect5},
we discuss limitations of our models and compare with related models.
Subsects. \ref{subsect4B} and \ref{subsect4C} present our results
with considerable detail. For readers not wishing this much detail, the 
first part of Sect. \ref{sect5} concisely summarizes the basic results
before proceeding to give our conclusions about them.

\section{Models}
\label{sect2}

This section reviews the kinematics of DPE dijet production and the
formulas for the F(IS)DPE and N(L)DPE models, based on the presentation in
\cite{Berera:1996vi}.  

\subsection{Kinematics}

\begin{figure}
   \begin{center}
      \leavevmode
      \epsfxsize=0.25\hsize
      \epsfbox{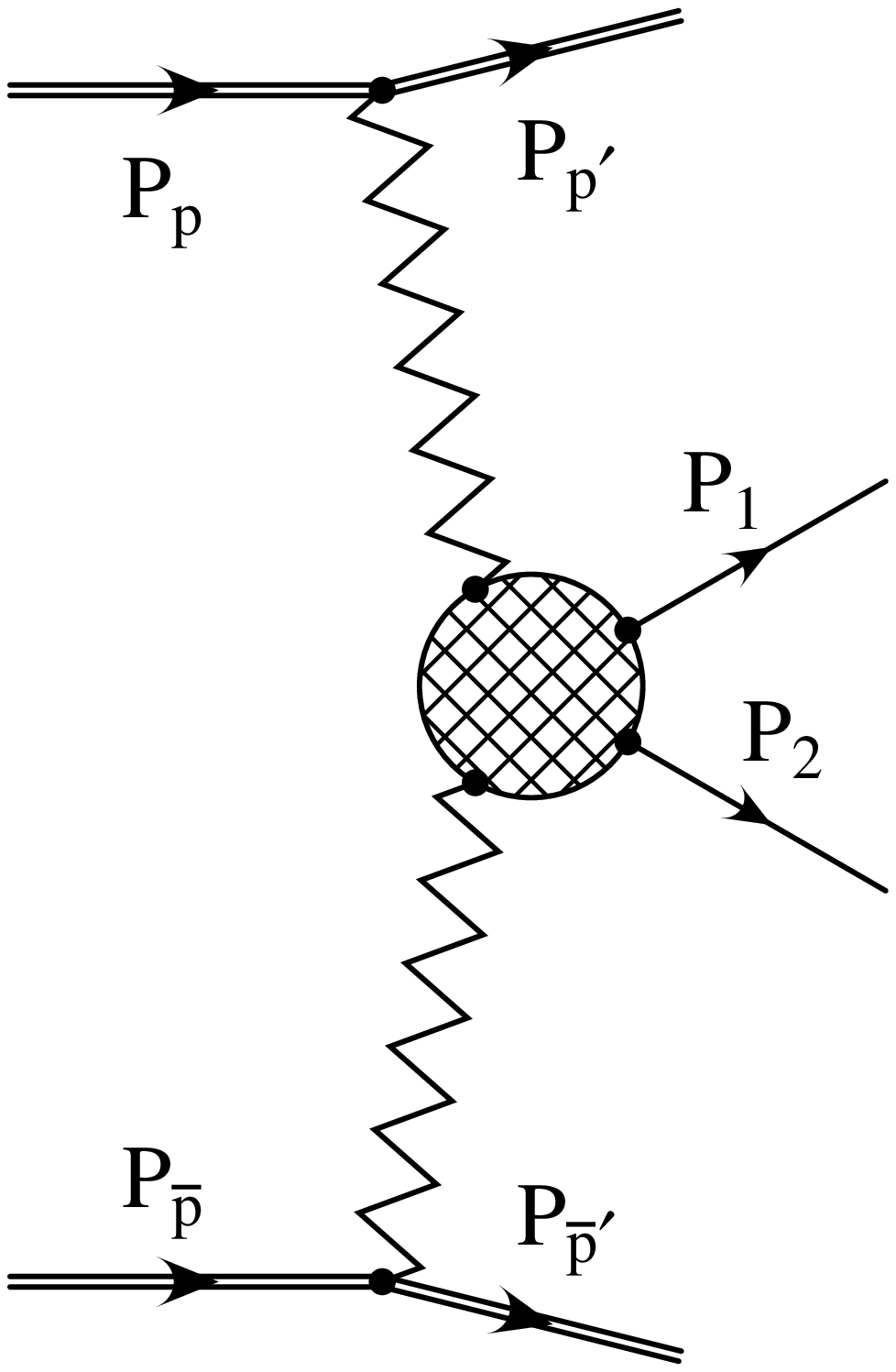}
   \end{center}
   \caption{
       Double pomeron exchange (DPE) to two jets.
   }
   \label{dpe2}
\end{figure}

The DPE dijet process examined in this paper is shown in 
Fig. \ref{dpe2},
\begin{equation}p+{\bar p}\rightarrow p'+{\bar p}'+\mbox{\rm 2 jets}.
\label{DPE.jets}
\end{equation}
The proton and antiproton
collide at high center of mass energy $s \equiv (P_p+P_{\bar p})^2 \rightarrow
\infty$, 
lose tiny fractions $x_{\pom/p}$ and $x_{\pom/{\bar p}}$ of their
respective longitudinal momenta,
and acquire transverse momenta ${\bf Q}_p$
and ${\bf Q}_{\bar p}$.  (This defines a diffractive regime, and in Regge
theory would lead to an expectation of the dominance of
double pomeron exchange (DPE)).
Using light-cone coordinates
$(+,-;{\bf \perp } )$,
the components of momenta of the hadrons
in Fig.~\ref {dpe2} are
\begin{eqnarray}
   P_{p}&=&\left(\sqrt {\frac {s}{2}},\frac {M^{2}}{\sqrt {2s}};{\bf 0}\right)
\nonumber\\
   P_{{\bar p}}&=&
\left(\frac {M^{2}}{\sqrt {2s}},\sqrt {\frac {s}{2}};{\bf 0}\right)
\nonumber\\
   P_{p'}&=& \left(
        (1-x_{\pom/p})\sqrt {\frac {s}{2}},
        \frac {(M^{2}+{\bf Q}_{p}^{2})}{ (1-x_{\pom/p})\sqrt {2s}};
        {\bf Q}_{p}
     \right)
\nonumber\\
P_{{\bar p}'}&=&
\left(\frac {(M^{2}+{\bf Q}_{\bar p}^{2})}
{(1-x_{\pom/{\bar p}})\sqrt {2s}},(1-
x_{\pom/{\bar p}})\sqrt {\frac {s}{2}};{\bf Q}_{\bar p}\right).\end{eqnarray}
Here we use bold-face type to indicate two-dimensional transverse
momentum.

The jets carry large momenta of magnitude
$E_T$ in the plane perpendicular to the collision axis
with azimuthal angle $\phi$.  (This defines a hard-scattering
regime.)
The small transfer of
longitudinal momentum to the hard process implies large rapidity
gaps between the jets and the two outgoing hadrons.  
The momentum delivered by the two incoming partons to the hard collision
that creates the jets is some portion $x_p$,$x_{\bar p}$ of the longitudinal
momentum fractions $x_{\pom/p}$, $x_{\pom/{\bar p}}$ respectively,
$0 < x_p \leq x_{\pom/p}$, $0 < x_{\bar p} \leq x_{\pom/{\bar p}}$.
Thus for the jets, 
ignoring terms of relative order $\ll 1$, the components of their
momenta are
\begin{eqnarray}
p_{1}&=&\left(ax_{p}\sqrt {\frac {s}{2}},bx_{\bar p}\sqrt {\frac
{s}{2}};E_{T}\cos\phi
,E_{T}\sin\phi \right)\nonumber\\
p_{2}&=&\left(bx_{p}\sqrt {\frac {s}{2}},ax_{\bar p}\sqrt {\frac
{s}{2}};-E_{T}\cos\phi
,-E_{T}\sin\phi \right),
\label{jetp}
\end{eqnarray}
where it is convenient to define
\begin{eqnarray}
a&\equiv &\frac {1+\sqrt {1-\kappa }}{2},\nonumber\\
b&\equiv &1-a,\end{eqnarray}
with
\begin{equation}
\kappa \equiv \frac {4E^{2}_{T}}{x_{p}x_{\bar p}s}.
\end{equation}
For later use, define the ratios
\begin{equation}
   \beta_{p}\equiv \frac {x_p}{x_{{\pom}/p}},
   ~\beta_{\bar p}\equiv 
    \frac {x_{\bar p}}{x_{{\pom}/{\bar p}}}.
\label{defbeta}
\end{equation}

It is conventional to describe the jet kinematics through the transverse
momentum $E_T$ in Eq. (\ref{jetp}) and the rapidity variables
\begin{equation}
   y_{1}\equiv \frac {1}{2}\ln\frac {p_{1}^{+}}{p_{1}^{-}},
   ~y_{2}\equiv \frac { 1}{2}\ln\frac {p_{2}^{+}}{p_{2}^{-}},
\end{equation}
which sometimes are expressed as
$y_+ \equiv (y_1+y_2)/2$ and $y_- \equiv y_1 - y_2$.
In terms of the jet rapidity variables and $E_T$, we have
\begin{eqnarray}
   x_{p}&=&\frac {E_{T}}{\sqrt {s}}(e^{y_{1}}+e^{y_{2}}),
\nonumber\\
   x_{\bar p}&=&\frac {E_{T}}{\sqrt {s}}(e^{-y_{1}}+e^{-y_{2}}),
\label{rapvar}
\end{eqnarray}
and
\begin{equation}
   \kappa =\frac {1}{\cosh^{2} (y_{-}/2)}.
\end{equation}

\subsection{Factorized (Ingleman-Schlein) DPE $-$ F(IS)DPE}

\begin{figure}
   \begin{center}
      \leavevmode
      \epsfxsize=0.25\hsize
      \epsfbox{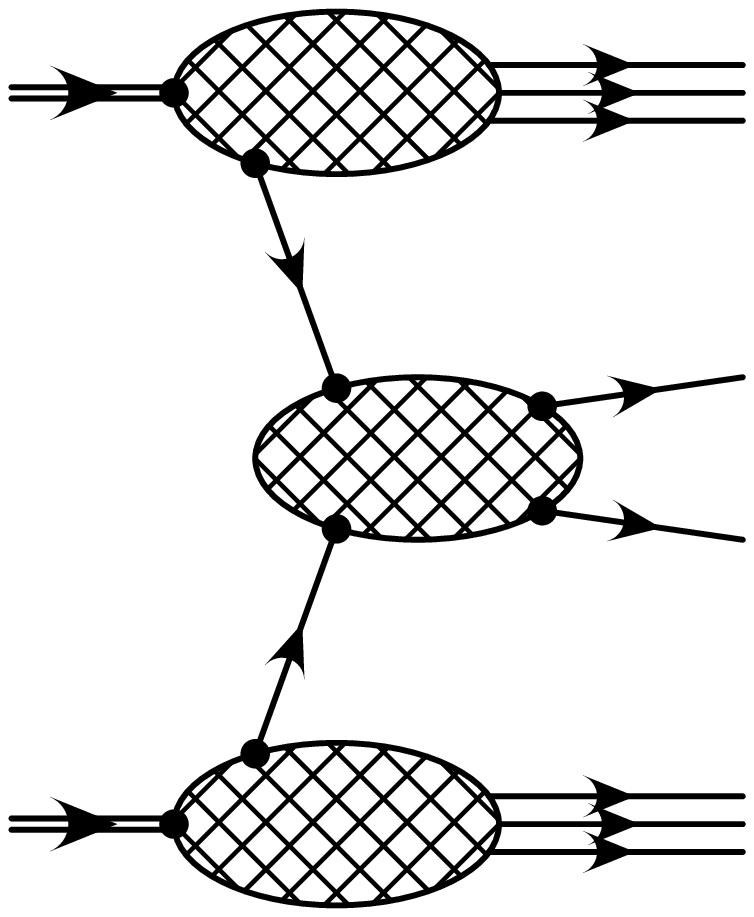}
   \end{center}
   \caption{
      Amplitude for inclusive two-jet production.
   }
   \label{dpincljets}
\end{figure}

To obtain the expression for the Factorized (Ingelman-Schlein)
DPE (F(IS)DPE) dijet differential cross section, first recall the
inclusive dijet differential cross section (Fig. \ref{dpincljets})
\begin{equation}
\frac{d \sigma^{dijet}_{incl.}}{dy_1 dy_2 dE_T^2} =
\frac{\pi}{s} \sum_{ij} [ f_{i/p}(x_p) f_{j/{\bar p}}(x_{\bar p}) 
{\hat \sigma}_{ij}({\hat s}, {\hat t}, {\hat u})
+ f_{j/p}(x_p) f_{i/{\bar p}}(x_{\bar p}) 
{\hat \sigma}_{ij}({\hat s}, {\hat t}, {\hat u})]/(1+\delta_{ij}),
\label{incjet}
\end{equation}
where $s$ is the CM energy between the two incoming hadrons (here protons),
$f_{i/h}(x_h)$,
is the inclusive parton distribution functions for
parton species $i$ in hadron $h$, and
$x_p, x_{\bar p}$ are the parton momentum fractions relative to proton
and antiproton respectively, carried by the two partons going
into the hard interaction.
${\hat \sigma}_{ij}$ is the parton 2 to 2 cross section for parton
species $i$ and $j$ with explicit expressions given
in \cite{ehlq}.

In the IS picture, they regard the pomeron as a hadronic particle.
The pomeron is hypothesized to be created 
from the incoming proton and
carries some momentum fraction $x_{{\pom}/h}$, $h=p,{\bar p}$,
of that proton's longitudinal
momentum.
In DPE hard expressions, one simply thinks of 
the collision of two pomerons in the same way as any two incoming hadronic
particles. As such, the inclusive dijet expression, Eq. (\ref{incjet}) above,
applies to this case with two modifications.
First $s$ now must be replaced by the appropriate CM energy for
the two pomerons, which is precisely
$x_{{\pom}/p} x_{{\pom}/{\bar p}} s$, where $s$ here is the CM energy
between the two incoming protons. Second a ``pomeron flux factor''
must be introduced, that expresses the probability to find a pomeron
inside the proton.

\begin{figure}
   \begin{center}
      \leavevmode
      \epsfxsize=0.25\hsize
      \epsfbox{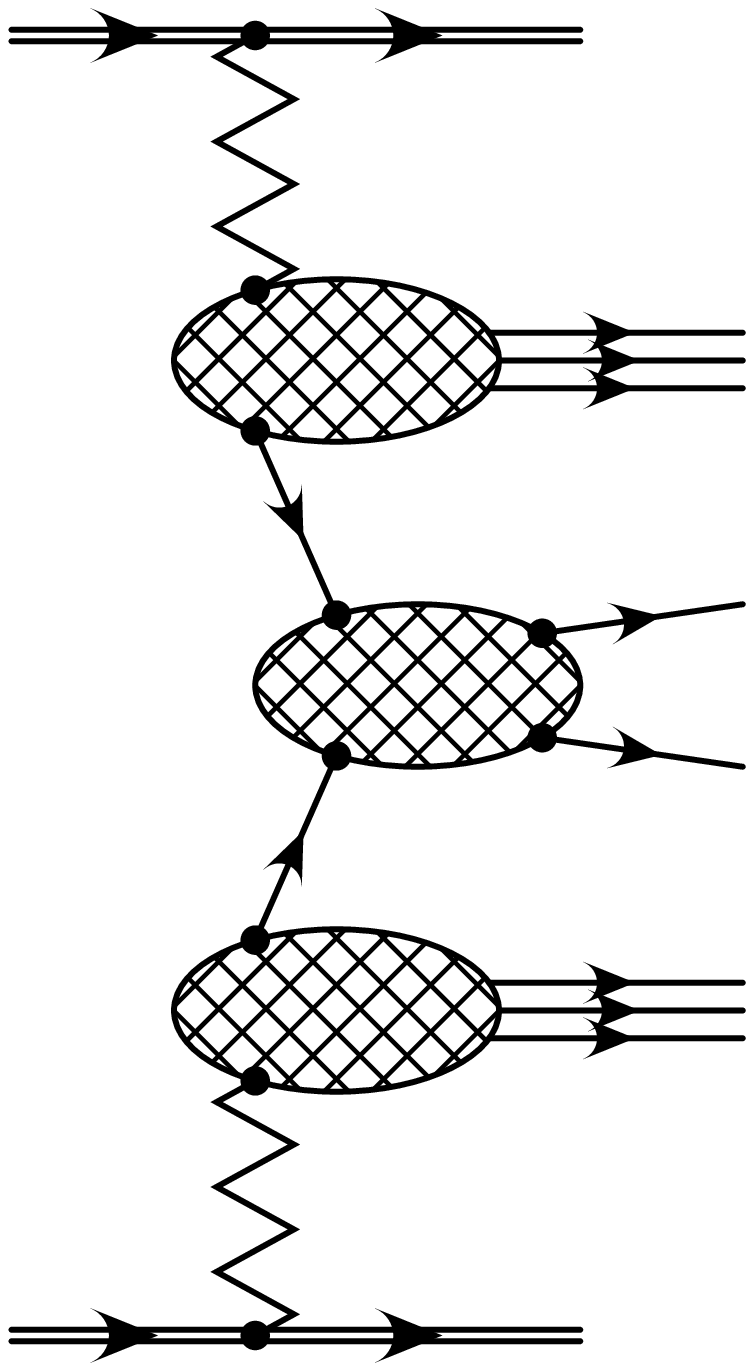}
   \end{center}
   \caption{
       Factorized (Ingelman-Schlein) 
       Double Pomeron Exchange (F(IS)DPE) amplitude
       with two jets produced.
   }
   \label{dpfdpe}
\end{figure}

With these considerations in mind,
the expression to F(IS)DPE dijet differential cross
section is (Fig. \ref{dpfdpe})
\begin{eqnarray}
\frac{d \sigma^{dijet}_{F(IS)DPE}}{dy_1 dy_2 dE_T^2} & = &
\int dx_{{\pom}/p} dx_{{\pom}/{\bar p}} f_{{\pom}/p}(x_{{\pom}/p})
f_{{\pom}/{\bar p}}(x_{{\pom}/{\bar p}})
\frac{\pi}{x_{{\pom}/p} x_{{\pom}/{\bar p}} s}
\nonumber \\
& & \sum_{ij} 
[ f_{i/\pom}(\beta_p) f_{j/\pom}(\beta_{\bar p}) 
{\hat \sigma}_{ij}({\hat s}, {\hat t}, {\hat u})
\nonumber \\
& + & f_{j/\pom}(\beta_p) f_{i/\pom}(\beta_{\bar p}) 
{\hat \sigma}_{ij}({\hat s}, {\hat t}, {\hat u})]/(1+\delta_{ij}).
\end{eqnarray}
In this expression 
$x_p, x_{\bar p}$ again are the momentum fractions of the incoming 
partons relative to the respective protons and
$\beta_h$ are 
the parton momentum fraction with
respect to the pomeron,
as defined in Eq. (\ref{defbeta}).
$f_{i/{\pom}}(\beta_h)$ now is the pomeron parton
distribution function.
${\hat \sigma}_{ij}$ is the 
parton 2 to 2 cross section, which is the same as
in the above inclusive case Eq. (\ref{incjet}).
Finally, $f_{{\pom}/h}(x_{{\pom}/h})$ is the
"pomeron flux factor". 
In our work, we will use the pomeron flux factor of Donnachie and
Landshoff\footnote{There is another commonly used pomeron flux factor
which is of Ingelman and Schlein \cite{Ingelman:1985ns}.
This differs from the DL flux factor primarily in its normalization.
However a change in the normalization factor completely is compensated for
by changing the parton densities by an inverse factor.  Thus the parton
densities are obtained, for example
in \cite{Alvero:1998ta}, for a set
of data without any {\it a priori} expectations as to their
normalization.} \cite{DL.norm},
\begin{equation}
f_{{\pom}/p}^{DL}(x_{\pom}) = 
f_{{\pom}/{\bar p}}^{DL}(x_{\pom}) = 
\int_{-1}^0 dt 
\frac{9 \beta_0^2}{4 \pi^2} 
\left[ \frac{4m_p^2-2.8t}{4m_p^2-t}
\left(\frac{1}{1-t/0.7}\right)^2 \right]^2
x_{\pom}^{1-2\alpha(t)},
\end{equation}
where $m_p \approx 0.938 {\rm GeV}$ is the proton mass, 
$\beta_0 \approx 1.8 {\rm GeV}^{-1}$ is the pomeron-quark coupling
and $\alpha(t)=\alpha_{\pom} + 0.25 t$ is the pomeron trajectory.
$\alpha_{\pom}$ is known as the pomeron intercept which for the soft
Pomeron is $\alpha_{\pom} \approx 1.08$ \cite{DL}.
The pomeron parton densities used here are those of
Alvero, Collins, Terron and Whitmore 
(hereafter ACTW) \cite{Alvero:1998ta}.
Their fits were to diffractive deep inelastic
and diffractive photoproduction of jets,
in which $\alpha_{\pom}$ was a free parameter that was fit to data
and found to be $\alpha_{\pom} \approx 1.14$.

The ACTW fits are to five models,
which covers a very
general set of possibilities.
Retaining their notation, the models will be denoted
as ACTW A, B, C, D, and SG. The precise description of these models can
be found in Sect. IID.  of their paper.  In brief,
the models A-D use conventional shapes for the initial distributions.
Model A represents a conventional hard quark parametrization, B has
in addition to A an initial gluon distribution, C 
has in addition to A a soft
quark distribution, and D has both additions to A. 
The final model, SG, has a gluon distribution that is peaked near
$\beta =1$. This form was motivated by the fit obtained by the H1
collaboration. In \cite{Alvero:1998ta}, they refer to it also as 
the "super-hard gluon".  

\subsection{Nonfactorized (Lossless) DPE $-$ N(L)DPE}

Our expression for the N(L)DPE dijet cross section is based on the toy
quantum field theory model in \cite{Berera:1996vi} which in effect
is the model of Low-Nussinov-Gunion-Soper \cite{lownus,gunsop}. 
The N(L)DPE dijet cross section
expression obtained here extends from \cite{Berera:1996vi} to account for 
the one-loop Sudakov
suppression factor.  We presented preliminary 
results with Sudakov suppression
in \cite{Berera:1997em}.  Our treatment of Sudakov suppression is the
same as by Martin, Ryskin, and Khoze\footnote{In 
\cite{Martin:1997kv} two types of double diffractive 
dijet expressions are given, which they
call exclusive and inclusive.  Both these expressions are nonfactorizing
processes of the CFS type, 
with the exclusive case the same as our N(L)DPE model.
Their inclusive case implements the same two gluon nonfactorizing
mechanism as their exclusive case.  The difference is, for the inclusive
case the incoming protons can diffract to any final state, provided
there are rapidity gaps between these final states and the hard
process.  This process is not relevant to this paper.}
\cite{Martin:1997kv}.  In fact, at one-loop order the nonabelian expression
required here is the same as the abelian expression 
of Sudakov \cite{Sudakov:1956sw} which in the context of
hard scattering was obtained
earlier by Collins \cite{Collins:1980ih}.  The only difference is, the abelian
expression must be multiplied by an overall group theory factor to
account for the additional color degrees in the nonabelian case.

\begin{figure}
   \begin{center}
      \leavevmode
      \epsfxsize=0.4\hsize
      \epsfbox{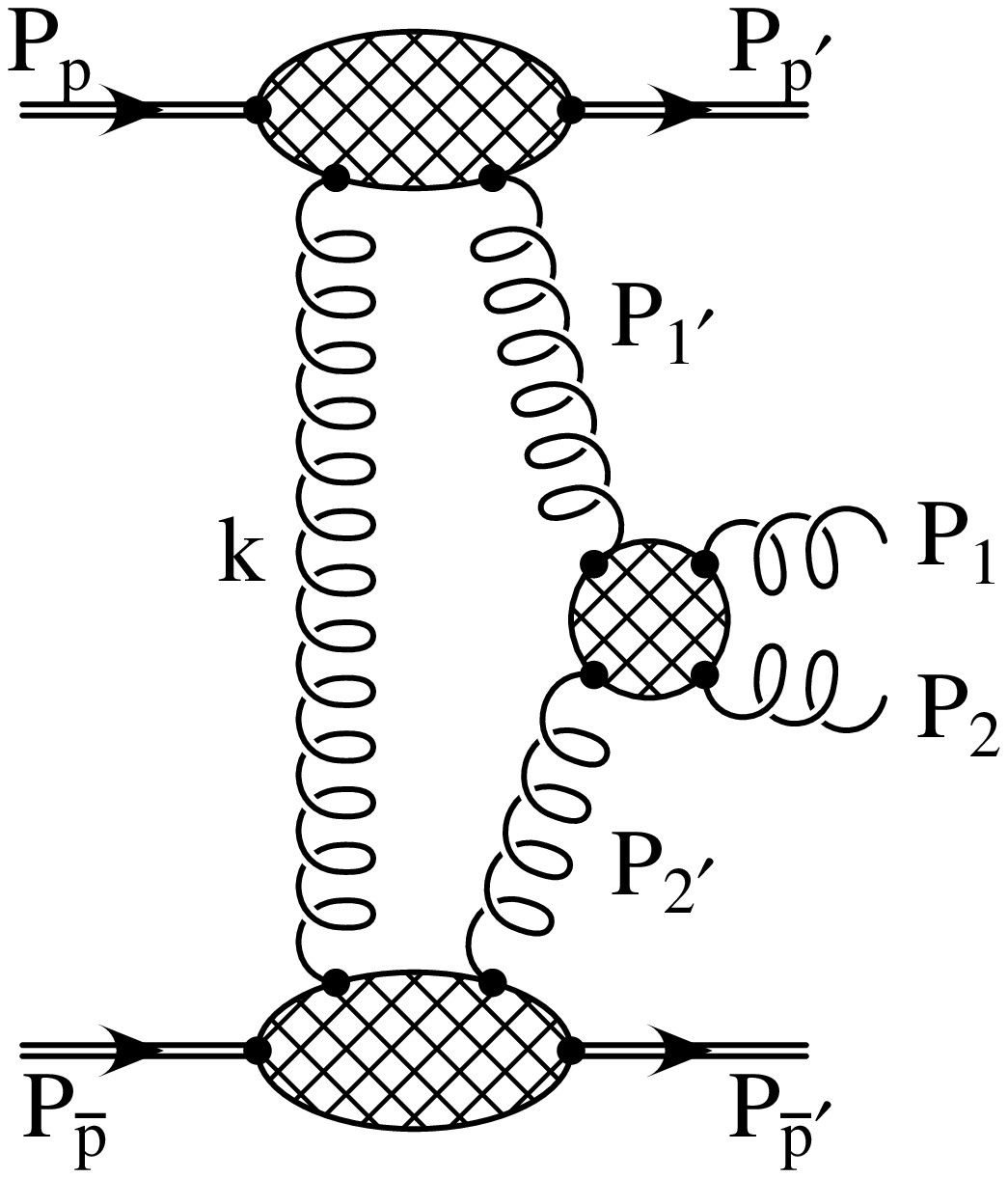}
   \end{center}
   \caption{
      Our model of the
      Nonfactorizing (Lossless) Double Pomeron Exchange (N(L)DPE) amplitude
      with two gluon jets produced.
   }
   \label{dpamp12}
\end{figure}

For the N(L)DPE model in Fig. \ref{dpamp12}b, 
$x_{\pom/p}$ and $x_{\pom/{\bar p}}$ again are
the longitudinal momentum fractions lost by proton and antiproton
respectively. In difference to the F(IS)DPE case, the momentum fractions
for the incoming partons to the hard process are equal to those lost by
the protons, $x_p=x_{\pom/p}$ and $x_{\bar p}= x_{\pom/{\bar p}}$
or equivalently $\beta_p=\beta_{\bar p}=1$.
Qualitatively this means all the momentum lost by the 
diffractive protons is transferred into the hard process.  This
kinematics is similar to the superhard component reported
for the case of single-sided diffractive dijet production
by the UA8 \cite{Brandt:1992zu}.

Our expression for the N(L)DPE dijet differential cross section is
\begin{equation}
   \frac {d\sigma^{dijet}_{N(L)DPE}}{dE^{2}_{T}
dy_{-}dy_{+}}
   = \int d^2{\bf Q}_p d^2{\bf Q}_{\bar p}
\frac {|\overline {{\cal M}}|^{2}\kappa ^{2}}{2^{16}\pi ^{ 7}E_{T}^{4}},
\label{dsig2}
\end{equation}
where
\begin{eqnarray}
   {\overline {\cal M}}
&=&-(-i)\int \frac {d^{2}{\bf k}}{(2\pi )^{2}}\,
   \hat {g}_{p}({\bf k},-{\bf Q}_{p})
   \,\hat {g}_{\bar p}({\bf k},{\bf Q}_{\bar p})\,
   \epsilon _{ i}({\bf k}{\bf -}{\bf Q}_{p})
   \,\epsilon _{j}(-{\bf k}{\bf -}{\bf Q}_{\bar p})
   \,{\cal A}(i,j;f) F_S({\bf k}^2,E_T^2).
\label{mdp}
\end{eqnarray}
Here, the ``polarization" vectors are defined as
\begin{equation}
   \epsilon _{i}({\bf k})=\frac {{\bf k}_{i}}{\sqrt {{\bf k}^{ 2}}},
\label{pol}
\end{equation}
${\hat g}_h({\bf k}, {\bf Q})$ are hadronic form factors with the explicit
expressions of our model in Eqs. (10) - (12) of \cite{Berera:1996vi}, and
${\cal A}(i,j;f)$ is the hard amplitude.  Two hard subprocesses are
possible, $g_{1^{\prime}} g_{2^{\prime}} \rightarrow g_1 g_2$ and
$g_{1^{\prime}} g_{2^{\prime}} \rightarrow q_1 {\bar q}_2$.
The calculations in \cite{pumplin,Berera:1996vi} showed
that the latter process
gives zero contribution to the N(L)DPE dijet cross section
when the final state transverse momentum of the two diffractive hadrons 
is zero.  This should suppress quark jet production relative
to gluon jet production.
In \cite{Berera:1996vi} this expectation was explicitly confirmed.
Thus only the $g_{1^{\prime}} g_{2^{\prime}} \rightarrow g_1 g_2$ 
hard process is relevant for N(L)DPE production..
The explicit expressions for ${\cal A}(i,j;f)$ for this
process are given in
Appendix A of \cite{Berera:1996vi}.  The Sudakov suppression factor
is $F_S({\bf k}^2,E_T^2)$, which at one-loop order 
is \cite{Collins:1980ih,Martin:1997kv}
\begin{equation}
F_S({\bf k}^2,E_T^2) = \exp\left[-S({\bf k}^2,E_T^2\right],
\label{sudsup}
\end{equation}
where 
\begin{equation}
S({\bf k}^2,E_T^2) = \frac{3 \alpha_S(E_T^2)}{4 \pi} \ln^2 
\left[\frac{E_T^2}{4({\bf k}^2 + \mu^2)}\right]
\end{equation}
and $\mu$ ($\stackrel{<}{\sim} 1 {\rm GeV}$) is a low energy cutoff scale.

The amplitude ${\overline {\cal M}}$ in Eq. (\ref{mdp}) is fixed up to
an overall normalization which implicitly is specified through the
hadronic form factors ${\hat g}_h$.  Based on the same quantum field theory
model, an expression for the hadron-hadron elastic scattering amplitude
can be determined and that expression involves the same hadronic form
factors ${\hat g}_h$. In fact this expression for the elastic scattering
amplitude essentially is the one of Low-Nussinov \cite{lownus}
and Gunion-Soper \cite{gunsop}. Thus the free parameter in our
N(L)DPE model that fixes the overall normalization of  
${\overline {\cal M}}$ is chosen to yield the experimental value of the
$p {\bar p}$ forward elastic cross section.  The details of this
procedure are given in \cite{Berera:1996vi}.

\section{Experimental Cuts}
\label{sect3}

This section reviews details about the CDF and D\O\ experiments that are
relevant to the DPE dijet process.  In particular, we state the cuts we
will use to represent the CDF and D\O\ DPE dijet 
experiments\footnote{Some years back the UA1 also had reported
on jet events with double rapidity gaps in ${\bar p} p$-collisions at 
$\sqrt{s}=630 {\rm GeV}$ \cite{Joyce:1993ru}.
However their reported
results are insufficient to include in our analysis.}. Also, cuts
are given that are representative cases for DPE dijets at LHC.

CDF has presented results on double diffractive dijet production
\cite{cdfdd1,cdfdd2,cdfdd3,cdfdd4} 
at $\sqrt{s}=1800 {\rm GeV}$ with transverse jet
energies $E_T > 7 {\rm GeV}$.  The experiment has one Roman pot on the
$-$ rapidity side, which detects the diffractive hadron, here
${\bar p}$, going in this direction with the cuts
$0.04 < x_{{\pom}/{\bar p}} < 0.095$.
On the + rapidity side, there is no Roman pot, only
a rapidity gap requirement. Thus, in principle, there is no specified
cuts on the outgoing diffractive hadron, here $p$, that goes in the +
rapidity side.  However, based on the rapidity gap length on this
side, they obtain the estimate 
$0.015 < x_{{\pom}/{p}} < 0.035$. The experiment places no explicit cuts
on the dijet rapidity region.  In our calculations, as our cuts,
we will use the entire central detector region
$-4.2 < y_1,y_2 < 2.0$.

The CDF double diffractive dijet experiment has three shortcomings in
its interpretation as the DPE dijet process.
First, the lack of a Roman pot on the + rapidity side
to detect the proton is a primary source of ambiguity in differentiating
rapidity gap events that involve diffractive excitation of the proton
versus pomeron exchange. Second, a heuristic guide for pomeron exchange
is that in the diffractive event 
$x_{{\pom}/p,{\bar p}} < 0.05$. Above this limit
other Regge exchanges may be important or the interpretation as a
diffractive process altogether may be questioned.  Based on this guide,
the CDF Roman pot cuts on the diffractive antiproton,
$0.04 < x_{\pom/{\bar p}} < 0.095$, exceed the optimal region for
interpretation as pomeron exchange.
Third, CDF does not implement on-line jet triggering. Instead, they
collect a sample of events that have a ${\bar p}$ track in the Roman pot
and a rapidity gap in the +rapidity, $p$, side. From these events, they
separate those cases in which there are two jets in the central region
with $E_T> 7 {\rm GeV}$ and discard the remaining data.  Since generally
jets are difficult objects to create and the central region typically is
soft, most of the collected data is discarded.  Furthermore,
$ E_T > 7 {\rm GeV}$ is a relatively low transverse jet momentum requirement
for Tevatron jets
and at this level jets are fairly cloudy objects
that may be difficult to reconstruct and measure precisely.

D\O\ has been examining double diffractive dijet production at two
center-of-mass energies $\sqrt{s}=630 {\rm GeV}$ and $1800 {\rm GeV}$
\cite{d0dd1,d0dd2,d0dd3,d0dd4}.
Hereafter these two cases will be referred to respectively as
D\O\-630 and D\O\-1800.  At present, the D\O\ experiment has no Roman
pots\footnote{In Run II, which is expected to start in
late 2000, D\O\ plans to have Roman pots on both sides
of the beamline to detect both the proton and antiproton \cite{abran}},
so that DPE dijet production operationally is defined as two hard
jets in the final state which are separated from both sides of the
beamline by large rapidity gaps.  The dijet cuts are
$E_T > 12 {\rm GeV}$ for D\O\-630 and $E_T > 15 {\rm GeV}$ 
for D\O\-1800
with $|y_1|,|y_2| < 1.0$ for both center-of-mass energies.

The D\O\ approach has both an advantage and a disadvantage to the CDF
approach. The advantage is D\O\ implements on-line jet triggering. Thus,
they are able to make a more efficient usage of their collected data
sample.  In addition 12 and 15 GeV jets are much better defined for
identification by cone algorithms.  The disadvantage of the D\O\ approach 
is without Roman pots the same ambiguities experienced by CDF arise
here and are magnified.  In particular uncertainty remains about what
portion of the double gap events are double diffractive, single
diffractive/single pomeron exchange or double pomeron exchange.
Moreover the momentum fractions lost by the proton and antiproton are
not directly measured, so that in principle the experiment places no
restriction on them.  An upper bound on the momentum fraction
lost by the $p$ and ${\bar p}$ can be estimated by examining the
maximum energy deposition in the hard event.  From this, it can be
inferred \cite{d0dd1} that
$x_{{\pom}/p}, x_{{\pom}/{\bar p}} \stackrel{<}{\sim} 0.05$.
This bound is consistent with the heuristic notion that a pomeron can
carry no more than $ \sim 5 \%$ of the diffractive hadron's momentum.
For the D\O\ cuts used in our calculations, we will use the upper
bound $x_{{\pom}/p}, x_{{\pom}/{\bar p}} < x_{\pom}^{\rm max} =0.05$.
The jet $E_T$ requirements imply a minimum energy must be deposited in
the hard interaction region, which places a lower bound 
bound $x_{{\pom}/p}, x_{{\pom}/{\bar p}} > x_{\pom}^{\rm min}$ given by
\begin{equation}
E_T^{\rm min} =
\sqrt{\frac{x_{\pom}^{\rm min} x_{\pom}^{\max} s}{2}}.
\label{xpmin}
\end{equation}
For convenience, to accommodate
both center-of-mass energy cases, 
we set $x_{\pom}^{\rm min} =0.001$,
since this limit is lower than the
values given  by Eq. (\ref{xpmin}).

For DPE dijets at LHC with $\sqrt{s}=14,000$ GeV, we use the following
cuts.  The transverse jet energy will be $E_T > 10 {\rm GeV}$
with the rapidity region $-1.0 < y_1,y_2 < 1.0$.  These cuts
represent standard expectations for jets in the hard DPE process.  For
the momentum fractions lost by the proton and antiproton, four cases
will be considered,
LHC-1: $0.002 < x_{{\pom}/p,{\bar p}} < 0.03$,
LHC-1': $0.00006 < x_{{\pom}/p,{\bar p}} < 0.03$,
LHC-2: $0.002 < x_{{\pom}/p,{\bar p}} < 0.01$, and
LHC-2': $0.00006 < x_{{\pom}/p,{\bar p}} < 0.01$.
These cuts are estimates of where diffraction should be important. The
upper bounds are slightly more conservative than the heuristic limit of
$0.05$.  For the LHC-2 and LHC-2' cuts, where the upper bounds are
$0.01$, diffraction clearly should be dominant as supported by the ZEUS
\cite{zeus} and H1 \cite{h1} experiments.
The lower bounds on $x_{{\pom}/p,{\bar p}}$
for LHC-1 and LHC-2 again are suggested by
the Zeus and H1 data.  The lower bound
on $x_{{\pom}/p,{\bar p}}$ for the LHC-1' and LHC-2' cases 
is based on the minimal energy condition for the hard interaction
Eq. (\ref{xpmin}).  Here, the lower limit 
of $x_{{\pom}/p,{\bar p}} > 0.00006$
accommodates both cases, LHC-1' and LHC-2', since this bound is below
$x_{\pom}^{\rm min}$ given by Eq. (\ref{xpmin}).

\subsection{Summary of the Experimental Cuts}
For convenience, below the cuts we use to represent the
various experiments are summarized.

\noindent
{\bf CDF}: $\sqrt{s}=1800 {\rm GeV}$, $E_T > 7 {\rm GeV}$,
$-4.2 < y_1, y_2 < 2.0$, $0.015 < x_{{\pom}/p} < 0.035$ 
(+rapidity side), $0.04 < x_{{\pom}/{\bar p}} < 0.095$
($-$ rapidity side).

\noindent
{\bf D\O\-1800}: $\sqrt{s}=1800 {\rm GeV}$, $E_T > 15 {\rm GeV}$,
$-1.0 < y_1, y_2 < 1.0$, 
$0.001 < x_{{\pom}/p}, x_{{\pom}/{\bar p}} < 0.05$ 

\noindent
{\bf D\O\-630}: $\sqrt{s}=630 {\rm GeV}$, $E_T > 12 {\rm GeV}$,
$-1.0 < y_1, y_2 < 1.0$, 
$0.001 < x_{{\pom}/p}, x_{{\pom}/{\bar p}} < 0.05$ 

\noindent
{\bf LHC-1}: $\sqrt{s}=14,000 {\rm GeV}$, $E_T > 10 {\rm GeV}$,
$-1.0 < y_1, y_2 < 1.0$, 
$0.002 < x_{{\pom}/p}, x_{{\pom}/{\bar p}} < 0.03$

\noindent
{\bf LHC-$1^{\prime}$}: $\sqrt{s}=14,000 {\rm GeV}$, $E_T > 10 {\rm GeV}$,
$-1.0 < y_1, y_2 < 1.0$, 
$0.00006 < x_{{\pom}/p}, x_{{\pom}/{\bar p}} < 0.03$
 
\noindent
{\bf LHC-2}: $\sqrt{s}=14,000 {\rm GeV}$, $E_T > 10 {\rm GeV}$,
$-1.0 < y_1, y_2 < 1.0$, 
$0.002 < x_{{\pom}/p}, x_{{\pom}/{\bar p}} < 0.01$

\noindent
{\bf LHC-$2^{\prime}$}: $\sqrt{s}=14,000 {\rm GeV}$, $E_T > 10 {\rm GeV}$,
$-1.0 < y_1, y_2 < 1.0$, 
$0.00006 < x_{{\pom}/p}, x_{{\pom}/{\bar p}} < 0.01$

\section{Results}
\label{sect4}

In this section, the results of our calculations are presented.
Then, various cross-checks and features of the results are discussed.
Finally a qualitative comparison is made of our results with the
preliminary results from the CDF and D\O\ experiments.

\subsection{Presentation}
\label{subsect4A}

Calculations have been performed of the $y_+$-spectra, $E_T$ spectra,
and total cross sections for the N(L)DPE and F(IS)DPE dijet processes with
the CDF, D\O\ and LHC cuts that were discussed in Sect. \ref{sect3}.
For comparison, the standard inclusive dijet 
process, Eq. (\ref{incjet}), also has
been computed with cuts comparable to the corresponding CDF, D\O\ and
LHC DPE cuts.  In particular, the corresponding inclusive dijet cuts we use
are for CDF: $\sqrt{s} = 1800 {\rm GeV}$, $E_T > 7 {\rm GeV}$,
$-4.2 < y_1,y_2 < 2.0$,
D\O\-1800: $\sqrt{s} = 1800 {\rm GeV}$, $E_T > 15 {\rm GeV}$,
$-1.0 < y_1,y_2 < 1.0$,
D\O\-630: $\sqrt{s} = 630 {\rm GeV}$, $E_T > 12 {\rm GeV}$,
$-1.0 < y_1,y_2 < 1.0$,
and LHC: $\sqrt{s} = 14,000 {\rm GeV}$, $E_T > 10 {\rm GeV}$,
$-1.0 < y_1,y_2 < 1.0$.

Our results for the $y_+$ and $E_T$ spectra are presented in 
Figs. \ref{figdpcdf} - \ref{figdplhc}
for respectively CDF, D\O\-1800, D\O\-630 and LHC.
In all four of these
figures, the (a) graphs contains the $y_+$-spectra for the N(L)DPE
and F(IS)DPE models, the (b) graphs contains the ratio
of $y_+$ spectra between the DPE and inclusive processes,
$(d\sigma_{DPE}/dy_+)/(d\sigma_{incl.}/dy_+)$,
the (c) graphs contain the $E_T$-spectra for the N(L)DPE and F(IS)DPE
models and the (d) graphs contain 
the ratio of the $E_T$ spectra between the DPE and inclusive processes,
$(d\sigma_{DPE}/dE_T)/(d\sigma_{incl.}/dE_T)$.
In the CDF and D\O\
figures, the solid curves represent the F(IS)DPE ACTW A-SG models
and the dashed curves represent the N(L)DPE model with
Sudakov suppression factor Eq. (\ref{sudsup}) none (i.e. $F_S=1$)
and $\mu^2=1.0$, $0.3$ ${\rm GeV}^2$.
The D\O\ figures also have
dashed-dotted and dotted curves, which represent for the F(IS)DPE
and N(L)DPE processes respectively some modified cuts.  The
specifics of these curves will be explained at the appropriate
time in the discussion that follows.
For the LHC cuts in Fig. (\ref{figdplhc}), the F(IS)DPE ACTW D model
is represented for LHC-1,2 by the solid curves and LHC-1',2'
by the dashed-dotted curves and
the N(L)DPE Sudakov suppressed $\mu^2=0.3$ model is represented
for LHC-1,2 by the dashed curves and LHC-1',2' by the
dotted curves.

The total cross sections for all the DPE
dijet cases are in Table 1.
For the corresponding inclusive dijet cases, 
the total cross sections are
\begin{eqnarray}
\sigma _{\rm incl}^{\rm CDF}&=&1.9~\mb,\nonumber\\
\sigma _{\rm incl}^{\rm D\O\-1800}&=&0.023~\mb,\nonumber\\
\sigma _{\rm incl}^{\rm D\O\-630}&=&0.013~\mb,\nonumber\\
\sigma _{\rm incl}^{\rm LHC}&=&0.71~\mb.\nonumber\\
\label{incsigma}
\end{eqnarray}

For the inclusive process, the CTEQ5 parton distribution functions
were used \cite{cteq5}.
For the F(IS)DPE case, we use the best fit 
value of the pomeron intercept found by ACTW \cite{Alvero:1998ta},
$\alpha_{\pom} = 1.144$.  Also,
the ACTW pomeron parton distributions were dependent on
$\alpha_{\pom}$ as a consequence of their fitting procedure, and
we have used the ones
at $\alpha_{\pom}=1.144$.
Both types of distribution functions are evolved with 
three flavors of quarks,
and in all calculations, we set $\Lambda_{QCD} = 0.271 {\rm GeV}$.

Further details about the ACTW pomeron parton distribution functions can
be found in their paper \cite{Alvero:1998ta}.
However some relevant facts about the ACTW results are reviewed
here in order to put our calculations in perspective
with their results.  Amongst the five models of pomeron parton
distribution functions considered by ACTW, they found
that the ones with high gluon content gave the best fit to the diffractive
DIS and diffractive photoproduction data.  In their notation
the high gluon models are B,D and SG, with D giving the best fit,
whereas the low gluon or quark dominated models are A and C.
As a cross check, the predictions of their
fitted models were examined in \cite{Alvero:1998dw} for charm production
in e-p collisions.  The cross sections for the high gluon models
were within an order of magnitude of both the ZEUS and H1 data, 
with model D again doing the best, whereas the cross 
sections predicted from
the low gluon models, A and C, were two orders of magnitude
below the experimental data.  Thus, hard factorization for
diffractive lepton-hadron scattering appears to be well supported by
the ACTW analysis.  However, in confronting their fitted
models to diffractive hadron-hadron scattering, a pronounced
inconsistency with data occurs. For DPE dijet production, 
the analysis in \cite{Alvero:1998dw} found that the high 
gluon model cross sections were 20-300 times larger 
than the CDF data, with
model D having the largest discrepancy, whereas the low
gluon models actually agreed within a factor two of data.
Other tests made by ACTW in \cite{Alvero:1998ta} also revealed similar
inconsistencies for hard factorization in diffractive
hadron-hadron processes.

At the moment there is no explanation for this breakdown in hard
factorization, and before any insight may be gained,
it appears the situation still is in search of more tests of the
data.  This paper provides several comparison tests between
theory and experiment with the primary aim to discriminate
between the two hard DPE mechanisms.  However, {\it en route}, these
tests also supplement the ACTW hard factorization analysis
and may provide additional insight into the problems uncovered
in their work.  In particular,  
the dijet distributions calculated in this paper provide
more detailed predictions from the basic models than just
total cross sections, with which to confront data.
Furthermore, total cross sections are calculated for several
experimental cuts, so that ratios amongst them can be tested to data.
These ratios can help test the validity of the absorptive 
correction models, which, as 
discussed in the Introduction, in generic Regge physics
inspired models
are believed to be fairly
independent of the hard kinematics.
Thus, for example, for the CDF and D\O\-1800 cases, the absorptive
correction effects from these models would give the same
overall correction factor and so should cancel out
in the ratio between the two experimental cross sections.

\subsection{Discussion and Cross-Checks}
\label{subsect4B}

This subsection highlights some interesting features in the results and
explains their underlying origins.  The F(IS)DPE and inclusive cross
section formulas for dijet production are well known in the literature.
We simply will quote where necessary properties about the various
quantities that enter in these expressions such as the parton
distribution functions, pomeron flux-factor and hard matrix element.
On the other hand, the N(L)DPE dijet cross section formula is less
familiar. In \cite{Berera:1996vi} it was noted that for forward scattering of
both hadrons and with no Sudakov suppression factor, $F_S=1$,
the square of the amplitude Eq. (\ref{mdp}) becomes
\begin{equation}
   |\overline {{\cal M}}(0,0)|^{2}
   = 64\pi \left(\frac {d\sigma  (0)}{dt}\right)_{\rm el}
     \delta _{ij}\delta _{kl}H_{ijkl},
\label{forcx}
\end{equation}
where $H_{iijj}$ is the square of the
hard parton amplitude.  Its exact expression is given in
\cite{Berera:1996vi}, which evaluates to be
$\sum_{ij} H_{iijj} = 18 [4\pi \alpha(E_T^2)]^2 (N_c^2-1) \cosh^4(y_-/2)$.
Using these expressions, Eq.(\ref{dsig2}) crudely can be approximated by the
following expression which can be evaluated upon inspection,
\begin{equation}
\frac{d \sigma^{dijet}_{N(L)DPE}}{dE_T^2 dy_1 dy_2} \approx
\left(\frac{d \sigma}{dt} \right)_{el}
\int_{\stackrel{\rm approximate}{\perp-{\rm region}}}
d^2{\bf Q}_p d^2{\bf Q}_{\bar p} \frac{\sum_{ij} H_{iijj}}
{2^{10} \pi^6 E_T^4 \cosh^4(y_-/2)} |F_S({\bf 0},E_T^2)|^2,
\label{approxn1}
\end{equation}
where the Sudakov suppression factor is approximated at the
${\bf k}=0$ point.  In this expression, the integral
of the two diffractive protons' outgoing transverse momentum phase space
$({\bf Q_p}, {\bf Q_{\bar p}})$ can be approximated as
\begin{equation}
\int_{\stackrel{\rm approximate}{\perp-{\rm region}}}
d^2{\bf Q}_p d^2{\bf Q}_{\bar p} \approx  \pi^2 |t|_{max}^2,
\end{equation}
where $|t|_{max}$ is a fixed parameter that represents the characteristic
transverse momentum cut-off for the diffractive protons.
With these approximations, and
$(d\sigma/dt)_{el} =201 {\rm mb}/{\rm GeV^2}$,
which is obtained from the optical theorem from the total cross section
$\sigma_{tot}^{{\bar p}p} =62 {\rm mb}$ \cite{DL},
Eq. (\ref{approxn1}) becomes
\begin{equation}
\frac{d \sigma^{dijet}_{N(L)DPE}}
{dE_T^2dy_1 dy_2} \approx (45.8 {\rm mb}/{\rm GeV^2}) 
|t|_{max}^2 \frac{\alpha^2(E_T^2)}{E_T^4}.
\label{approxn2}
\end{equation}
To determine $|t|_{max}$, the above expression can be compared to the
exact numerical expression, Eq. (\ref{dsig2}), at one point. 
From this, we will set
$|t|_{max} \approx 0.26 {\rm GeV}^2$.

The three subsections to follow examine the CDF, D\O\ and LHC
cases in turn, with cross checks and explanations offered for the
various features of the results in 
Figs. \ref{figdpcdf} - \ref{figdplhc}.  One immediate cross check
of our results is the magnitudes of the cross sections. For the F(IS)DPE
case, we have verified that our results agree with \cite{Alvero:1998dw}.
For the N(L)DPE case\footnote{The exclusive double
diffractive model in \cite{Martin:1997kv} has a more detailed 
description of the two-gluon pomeron process compared to our model.
For this reason, direct comparison of total cross sections is not
possible between out model and theirs.  In the Conclusion, we will discuss
further the model in \cite{Martin:1997kv}}, 
it will be seen below that the exact numerical
results are consistent with the approximate expression Eq.
(\ref{approxn2}).

\subsubsection{CDF}

\begin{figure}
   \begin{center}
      \leavevmode
      \epsfxsize=1.0\hsize
      \epsfbox{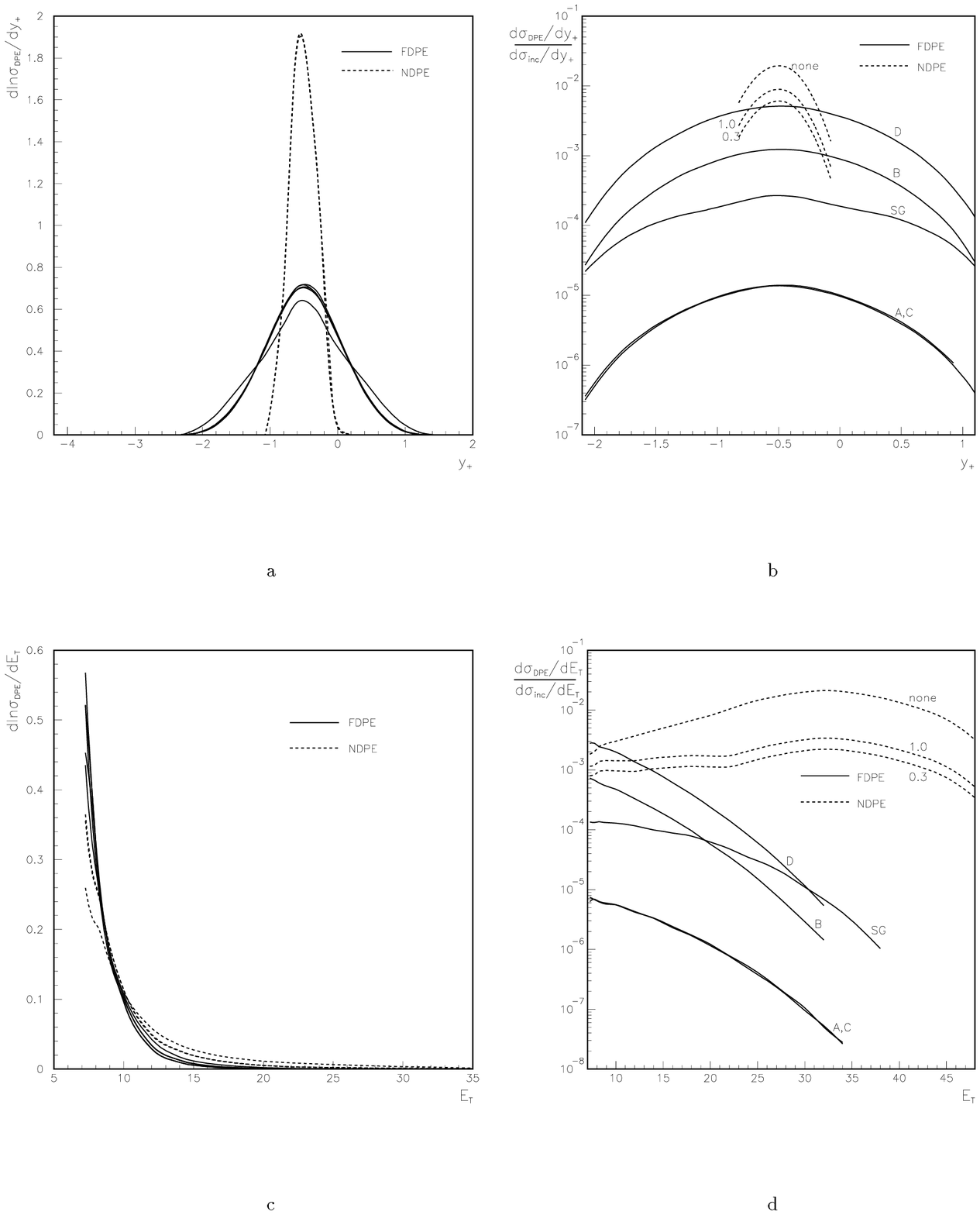}
   \end{center}
   \caption{
       Double pomeron exchange (DPE) dijet production-CDF cuts. (a) Mean
rapidity spectra $y_+ \equiv (y_1+y_2)/2$.
(b) Ratio of mean rapidity spectra between the DPE and corresponding
inclusive dijet processes.
(c) $E_T$-
Spectra. (d) Ratio of $E_T$-Spectra between
the DPE and corresponding
inclusive dijet processes.
   }
   \label{figdpcdf}
\end{figure}

The CDF results in Fig. \ref{figdpcdf} 
have the following noteworthy features. From
Figs. \ref{figdpcdf}a and \ref{figdpcdf}b, 
the $y_+$-spectra for the N(L)DPE process (dashed curves) are localized
to the region $-1 \stackrel{<}{\sim} y_+ \stackrel{<}{\sim} 0$.
The $y_+$-spectra is much
broader for the F(IS)DPE (solid curves) versus N(L)DPE case, 
and from Fig. \ref{figdpcdf}b both are
less broad than the inclusive $y_+$-spectra. 
This difference in the broadness of the $y_+$-spectra
between the F(IS)DPE and N(L)DPE processes with
CDF cuts is one of the most pronounced signatures
found in this study that could help to differentiate
the two processes.  As will be seen below,
this difference reflects upon intrinsic kinematic differences
between the two processes, and thus is a reasonably model
independent feature.   
Turning to the $E_T$-spectra,
from Fig. \ref{figdpcdf}c the N(L)DPE 
process falls much slower than the F(IS)DPE
process.  In fact from Fig. \ref{figdpcdf}d, 
the N(L)DPE process is seen to be almost
flat for $E_T < 45 {\rm GeV}$ for the two cases with Sudakov suppression
and slightly rising for the case with no Sudakov suppression.
Then for $E_T > ~ 45 {\rm GeV}$, all three cases rapidly fall to zero.
For reference, an exactly flat spectra in 
Fig. \ref{figdpcdf}d would imply 
it has the same shape as the inclusive $E_T$-spectra.
Thus the two Sudakov suppressed N(L)DPE $E_T$-spectra have approximately
the same shape as the inclusive $E_T$-spectra.  
On the other hand, for the F(IS)DPE spectra, all five cases fall much more
rapidly in Fig. \ref{figdpcdf}d relative to the inclusive spectra, with
the SG case falling the least rapidly.

The behavior of the CDF N(L)DPE spectra can be understood from the
approximate Eq. (\ref{approxn2}) and by examining the dijet rapidity
phase space. Recall for N(L)DPE processes, the parton momentum
fractions $x_h$ that enter the hard interaction equal the corresponding
pomeron momentum fractions $x_{{\pom}/h}$,
$\beta_h \equiv x_h/x_{{\pom}/h}=1$.  
As such from Eq. (\ref{rapvar}), the cuts on
$x_{{\pom}/h}$ imply direct restrictions on the jet rapidities
$y_1$, $y_2$.  One finds upon inspection of Eq. (\ref{rapvar})
and the explicit CDF rapidity cuts from Sect. \ref{sect3}
that at $E_T=7 {\rm GeV}$, dijets
only appear in the rapidity ranges $1.4 \stackrel{<}{\sim} y_1 \leq 2$,
$-3.3 \stackrel{<}{\sim} y_2 \stackrel{<}{\sim} -2.3$ 
(and interchange $y_1 \longleftrightarrow y_2$),
which equivalently implies 
$-1.4 \stackrel{<}{\sim} y_+ \stackrel{<}{\sim} -0.15$.  As $E_T$ increases,
the kinematically allowed rapidity bands move inwards towards zero
rapidity. Generally both bands also get narrower. However, since the
rapidity band at the proton side (+rapidity) was prematurely cut-off at
$2$ due to the explicit rapidity cuts, this band first
broadens up to $E_T \sim 14 {\rm GeV}$ and then narrows thereafter
for higher $E_T$.  As such at $E_T=14 (28) {\rm GeV}$
dijets appear in the bands
$0.5 (-0.3) \stackrel{<}{\sim} y_1 \stackrel{<}{\sim} 1.5 (0.75)$, 
$-2.6 (-1.8) \stackrel{<}{\sim} y_2 \stackrel{<}{\sim} -1.5 (-0.7)$
(and interchange $y_1 \longleftrightarrow y_2$),
which corresponds to 
$-1.1 \stackrel{<}{\sim} y_+ \stackrel{<}{\sim} 0$ at both 
$E_T$ scales. These considerations suffice to explain   
the localized $y_+$-spectra in Figs. \ref{figdpcdf}a and \ref{figdpcdf}b 
for the N(L)DPE process.

Applying these estimates to Eq. (\ref{approxn2}), the differential
cross section $d\sigma_{N(L)DPE}/dE_T$ at, for example,
$E_T=7,14,28 {\rm GeV}$ gives the relative magnitudes
$1, 0.2, 0.03$, which are within a factor 2 of 
the exact numerical results in Fig. \ref{figdpcdf}c.
This figure also indicates that the region above
$E_T > ~ 30 {\rm GeV}$ accounts for less than 
$0.5\%$ of the total N(L)DPE cross section.  A check of the dijet
phase space indicates that above this $E_T$,
the accessible region is rapidly
diminishing.  In fact due to the kinematic constraints, 
the maximum energy that can be
deposited in the hard region for either the F(IS)DPE or N(L)DPE processes is
$\sqrt{x_{{\pom}/p}^{\rm max} x_{{\pom}/{\bar p}}^{\rm max} s}$
which for the CDF 
cuts implies the largest dijet $E_T$ is
$E_T \stackrel{<}{\sim} 51 {\rm GeV}$. 
For the N(L)DPE case, this cutoff is best seen in
Fig. \ref{figdpcdf}d. 
  
The last point to address about the CDF N(L)DPE process is the magnitude of
the total cross section. The exact numerical results are given in
Table 1.  Estimates based on Eq. (\ref{approxn2}), where the
phase space integral and all other quantities are approximated at 
$E_T = E_T^{\rm min}  =7 {\rm GeV}$, agree up to a factor 2 
with the results in
Table 1, including the ratio amongst the three
cases of Sudakov suppression, none, $\mu^2=1 {\rm GeV}^2$,
and $\mu^2 = 0.3 {\rm GeV}^2$, of respectively $1, \sim 0.5, \sim 0.25$.

Turing to the F(IS)DPE process, the first point to be
addressed is the steeper decrease of the $E_T$-
spectra in Figs. \ref{figdpcdf}c and \ref{figdpcdf}d 
relative to both the N(L)DPE and inclusive
$E_T$-spectra. Two facts are useful for this analysis.
First, as $E_T$ increases, in general, the average value
of $x_p$ and $x_{\bar p}$ increase, since more energy
must be deposited into the hard region.  Second, the pomeron parton
distribution functions
$f_{i/{\pom}}(\beta_h)$ at small argument 
$\beta_h \equiv x_h/x_{{\pom}/h}$ grow as
$f_{i/{\pom}}(\beta \rightarrow 0) \sim \beta^{-a}$ 
with $0 \stackrel{<}{\sim} a \stackrel{<}{\sim} 1.5$,
and at large argument vanish as
$f_{i/{\pom}}(\beta \rightarrow 1) \sim (1-\beta)^{b}$ 
with $0 \stackrel{<}{\sim} b \stackrel{<}{\sim} 1.0$.
Thus at small $E_T$, $x_p$ and $x_{\bar p}$, and so therefore also
$x_p/x_{{\pom}/p}$ and $x_{\bar p}/x_{{\pom}/{\bar p}}$, are closer
to their kinematic lower bounds, which implies
the parton distribution functions are at their largest.
However as $E_T$ increases, it implies
$\beta_h=x_h/x_{{\pom}/h} \rightarrow 1$ so that
$f_{i/{\pom}}(x_h/x_{{\pom}/h}) \rightarrow 0$.
In contrast, within the same $E_T$ range, the behavior
of the inclusive parton distribution functions is
very different, primarily due to the difference in behavior
of their arguments. The inclusive distribution is evaluated with respect
to $x_h$ not $x_h/x_{{\pom}/h}$. Since
$x_h^{\rm min} < x_h < x_{{\pom}/h}$ and $x_{{\pom}/h} < ~ 0.1$,
$x_h$ within this range always is relatively small. 
Thus the inclusive parton distributions 
within the equivalent $E_T$ range have less variation and generally are
large.
This difference in behavior of the arguments for
inclusive and F(IS)DPE parton densities
explains the steeper decline of the latter's $E_T$-spectra relative to the
former.

There are two immediate checks that verify the above observations
about the $E_T$-spectra. First
pomeron parton distribution functions that fall slower as
$\beta \rightarrow 1$ should have flatter $E_T$-spectra 
in Fig. \ref{figdpcdf}d
and this is the case for the ACTW SG model.  Second, if
the upper limits on $x_{{\pom}/p, {\bar p}}$ are increased, then
for fixed $x_h$, the ratio
$x_h/x_{{\pom}/h}$ is smaller.  Thus, for the same jet kinematics,
$f_{i/{\pom}}(x_h/x_{{\pom}/h})$ should fall less rapidly,
which in turn would flatten the $E_T$-spectra in Fig. \ref{figdpcdf}d.
One can verify this effect for any general pomeron parton
distribution function.

To further quantify the above observations about the CDF case, we can ask 
how small do the
the parton momentum fractions become.  From
Eq. (\ref{xpmin}), naturally the minimum value for one occurs, when the other
is at its maximum.  As a more realistic
estimate, let us assume the "large" region for the 
parton momentum fractions is when 
$x_h/x_{{\pom}/h} \stackrel{>}{\sim} 0.5$,
since above this point the pomeron parton distribution functions rapidly
vanish.  Thus the parton carrying the "large" momentum fraction will
have $x_{h_{large}} \approx 0.5 x_{{\pom}/h_{large}}^{max}$.
We substitute $x_{h_{large}}$ in the LHS of Eq. (\ref{xpmin}).
We now ask above what $E_T$ will the other parton momentum fraction also
be in the "large" region, under the assumption that when
both parton's momentum fractions are "large", there is negligible
contribution to the cross section.  By this criteria, we find 
that both momentum fractions are in the 'large" region, which means
$x_p \stackrel{>}{\sim} 0.025$ on the proton side and 
$x_{\bar p} \stackrel{>}{\sim} 0.03$ on the
antiproton side, once $E_T \stackrel{>}{\sim} 35 {\rm GeV}$.

Although for the pomeron case, these momentum fractions are large 
because $\beta_h \sim 1$, the situation is different for
the inclusive case.  The inclusive parton distribution functions are
evaluated with respect to $x_h$, not $\beta_h = x_h/x_{{\pom}/h}$.
As such the range for their arguments is
$x_{p,{\bar p}} \sim 0.02 - 0.04$ and within this 
range the parton distribution
functions have very little variation.
The numbers quoted in this example are very crude, but they illustrate the
reason in Fig. \ref{figdpcdf}d for 
the F(IS)DPE $E_T$ spectra's steeper decline relative
to the inclusive case. 
 
The above discussion ignored entirely complications from the pomeron flux
factor $f^{DL}$. This is because its approximate behavior is
$f^{DL} \sim x_{\pom}^{1-2\alpha_{\pom}}$ and in either of the two ranges
$0.015 < x_{{\pom}/p} < 0.035$ or 
$0.04 < x_{{\pom}/{\bar p}} < 0.095$, its variation is relatively small, 
i.e. less than a factor 3.

It is worth noting that the rise
in the N(L)DPE $E_T$ spectra in Fig. \ref{figdpcdf}d
has similar explanation to the one given above for
the differences in
$E_T$-spectra between the F(IS)DPE and inclusive cases.
In particular, the inclusive parton distribution
functions will fall a little as $E_T$ increases since the average values
of $x_h$ will increase.  However the proton form factors
in our N(L)DPE model are insensitive to this
variation.  This is one of the notable differences between the
N(L)DPE model and both the inclusive and F(IS)DPE models, and it explains
the relative rise in the former's $E_T$ spectra to the latter.
Furthermore, the rise is less pronounced for the Sudakov suppressed
N(L)DPE processes, since they provide greater
suppression to the N(L)DPE differential cross section
$d\sigma/dE_T$ as $E_T$ rises. 

In the N(L)DPE model,
this lack of $x_h$-dependence in the proton form factors
is not a fundamental requirement for nonfactorization
of the CFS type.  This is a simplying limitation in this
particular model.  In \cite{Martin:1997kv} nonfactorizing models
similar to our N(L)DPE model were treated, except with
more detailed modeling of the $x_h$ dependences.
We will discuss the models in \cite{Martin:1997kv} later in
the paper, but we will not explore such
modification to our model in this
paper.

Next, we will understand the behavior of the $y_+$-spectra for the F(IS)DPE
case in Fig. \ref{figdpcdf}a and \ref{figdpcdf}b.  
To simplify the problem, we assume the main
features of the spectra are determined by the low $E_T$ dijets
$E_T \approx E_T^{\rm min} =7 {\rm GeV}$. We want to understand why 
the $y_+$-spectra is much broader for the F(IS)DPE versus 
N(L)DPE case and moreover why the former also
is skewed towards the $- y_+$ side.
For this, note that at fixed $x_p$, $x_{\bar p}$, and $E_T$, the largest $y_+$
attainable by the dijets is when both are on the same side with equal
rapidity, $y_-=0$.  In this case
Eq. (\ref{rapvar}) becomes 
\begin{equation}
x_{\stackrel{p}{\bar p}}=\frac{2E_T}{\sqrt{s}} \exp(\pm y_+).
\end{equation}
By evaluating $x_{p,{\bar p}}$ within their allowed range,
this expression gives the limits on $y_+$.
The allowed ranges for $x_{p,{\bar p}}$ are determined by the same
criteria as before that
both momentum fractions must be
"small", $x_h/x_{{\pom}/h} < ~ 0.5$.  This condition implies
the ranges $x_p \stackrel{<}{\sim} 0.02$, 
$x_{\bar p} \stackrel{<}{\sim} 0.05$, with the lower limits in both
cases governed by the energetic condition Eq. (\ref{xpmin}).
By these crude approximations, the $y_+$ range for the
spectra is
$-1.9 \stackrel{<}{\sim} y_+ \stackrel{<}{\sim} 1.0$, 
which coincides reasonably well
with the exact numerical results in Fig. \ref{figdpcdf}a.  
Furthermore,
one finds as the boundaries of the
$y_+$ range are approached, $x_p$ and $x_{\bar p}$ are increasing
for both the inclusive and F(IS)DPE cases.  As such,
the contributions from the respective parton distribution functions are
decreasing.  In particular, the distribution functions decrease slower
for the inclusive versus F(IS)DPE case, since the range of the argument
in the former is much smaller $\stackrel{<}{\sim} 0.1$ versus the latter
$\stackrel{<}{\sim} 1$.  This part of the explanation is the same as our earlier
discussion which compared the $E_T$-spectra for
these two processes. The final outcome is the F(IS)DPE $y_+$-spectra 
for all the models in Fig. \ref{figdpcdf}b
are narrower than the inclusive one.

The last point to note is that the total cross sections in Table I for
the five models come in the ratio 1(A):100(B):1(C):400(D):20(SG).
To obtain insight into these ratios, it is useful to decompose the cross
section in terms of the parton initiated processes.  For the
B,D and SG models we find $\sim 80\%$ of the cross section comes from
the pure gluon process ${\hat \sigma}_{gg}$ and the remaining fraction
predominately from the ${\hat \sigma}_{gq}$ process.  On the other
hand, for the A and C models, the cross section decomposes as
less than $5\%$ from ${\hat \sigma}_{gg}$, 
$\sim 40\%$ from ${\hat \sigma}_{gq}$ and 
$\sim 60\%$ from the pure quark initiated process
${\hat \sigma}_{qq} + {\hat \sigma}_{{\bar q}q}$.
Therefore the B,D, and SG
models are more gluon controlled whereas the A and C models are more
quark controlled.  However,
for none of the five models is it the case that one species of
partons, quarks or gluons, dominates the cross section.

To cross check these findings, we examine the pomeron parton
distribution functions in the most probable $\beta$ range,
which we estimated above to be $\sim 0.01-0.1$. Two features about the
parton distribution functions are evident.  First, in this $\beta$
range, the A and C or B and D distributions are the same magnitude, with
the latter pair about a factor 10 larger than the former pair and the
SG distribution is a factor 3-5 larger than the former.
Second, for a given parton model, the ratio of the gluon to quark parton
distribution function in this $\beta$ range,
$f_{g/\pom}(\beta)/f_{q/{\pom}}(\beta)$, is for the
A and C models $3-20$, the B and D models $20-50$ and 
the SG model $10-30$.
These properties are consistent with the general trends found amongst
the five models for the total cross sections
in Table 1.  However
since this turns out to be an intermediate regime between quarks and
gluons, these simple indicators are insufficient to better quantify the
results found from the exact numerical calculations.

\subsubsection{D\O\ }

\begin{figure}
   \begin{center}
      \leavevmode
      \epsfxsize=1.0\hsize
      \epsfbox{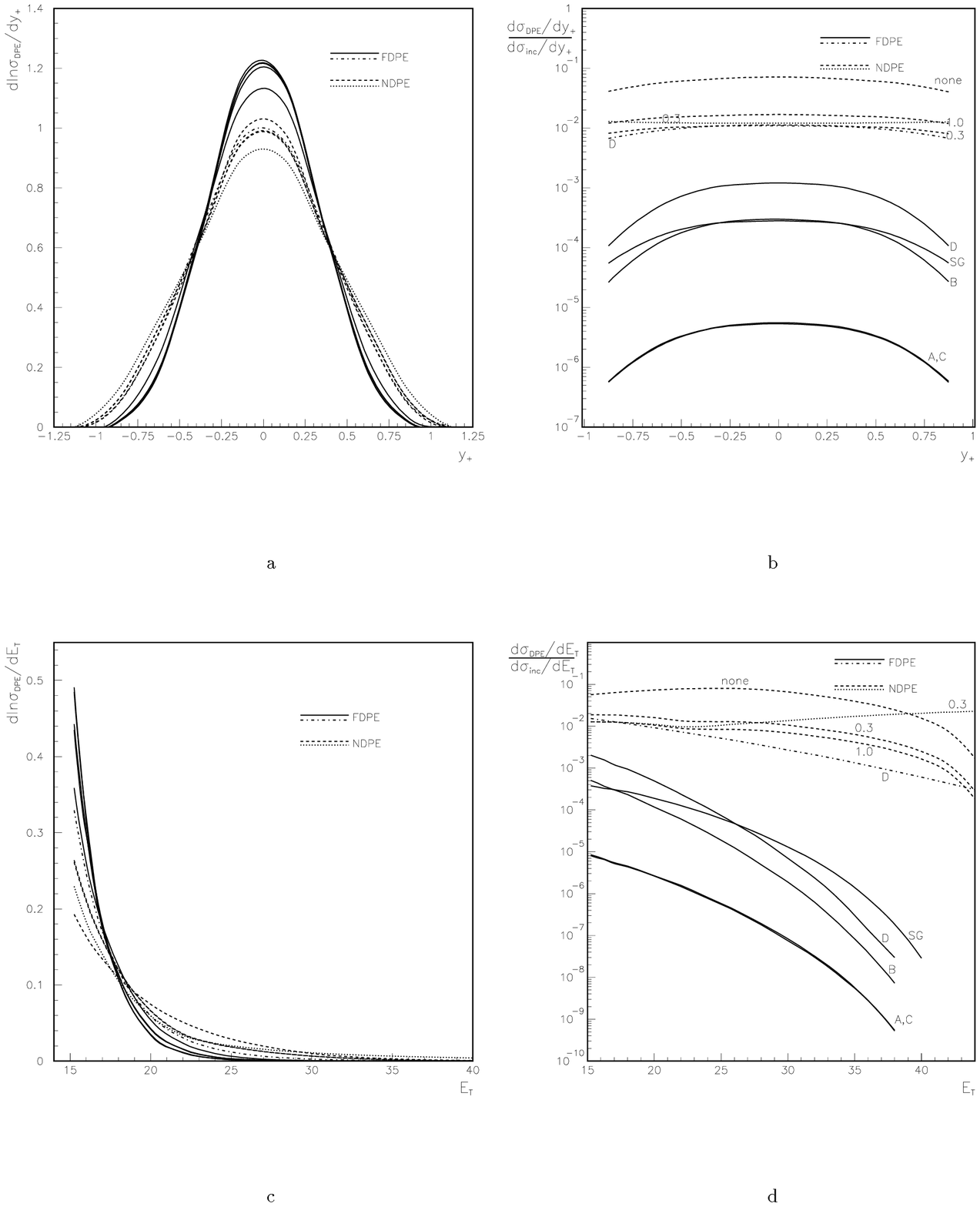}
   \end{center}
   \caption{
       Double pomeron exchange (DPE) dijet production - 
D\O\-1800 cuts. (a) Mean
rapidity spectra $y_+ \equiv (y_1+y_2)/2$.
(b) Ratio of mean rapidity spectra between the DPE and corresponding
inclusive dijet processes.
(c) $E_T$-
Spectra. (d) Ratio of $E_T$-Spectra between
the DPE and corresponding
inclusive dijet processes.
   }
   \label{figdpd0}
\end{figure}

The D\O\-1800 results in Fig. \ref{figdpd0} 
have the following interesting features. 
Both the N(L)DPE (dashed curves) and 
F(IS)DPE (solid curves) $y_+$-spectra are localized to the region 
$|y_+| < 1$. In contrast to the CDF case, here the F(IS)DPE $y_+$-spectra is
slightly narrower than the N(L)DPE $y_+$-spectra, which best is seen
in Fig. \ref{figdpd0}b.  For the $E_T$-spectra,
similar to the the CDF case the N(L)DPE process falls 
with increasing $E_T$ much slower
than the F(IS)DPE process.  In comparison to the inclusive process, for the
N(L)DPE case both the $y_+$-spectra in Fig. \ref{figdpd0}b 
and $E_T$ spectra for 
$E_T \stackrel{<}{\sim} 40 {\rm GeV}$ 
in Fig. \ref{figdpd0}d are flat, thus have the same shape
as the corresponding inclusive spectra.  On the other hand, for the F(IS)DPE
case, the $y_+$-spectra in Fig. \ref{figdpd0}b 
is much more localized than the
inclusive spectra and the $E_T$-spectra 
in Fig. \ref{figdpd0}d falls much faster.

The basic features of the N(L)DPE spectra can be understood, once again,
through the approximate cross section formula Eq. (\ref{approxn2})
and by examining the available dijet rapidity region based on the
explicit D\O\ rapidity cuts and those implied by
the cuts on $x_{{\pom}/p,{\bar p}}$.  Carrying out this analysis,
at the lowest $E_T=15 {\rm GeV}$ we find that the cuts on
$x_{{\pom}/p,{\bar p}}$ place no additional restrictions, so that the
available rapidity region is $-1 < y_1, y_2 < 1$.
As $E_T$ increases, the first rapidity region that diminishes is
for same-side jets and starting at the periphery.  The reason is
evident from Eq. (\ref{rapvar}).   At fixed $|y_1|$ and $|y_2|$,
$x_p$ and $x_{\bar p}$ differ more for same-side dijets than for
opposite-side dijets.  As such, the larger of the two parton momentum
fractions will reach its upper limit at
smaller $y_1$ and $y_2$ for same-side versus opposite-side
dijets. 
For example, the available rapidity region at 
$E_T =20 (30) {\rm GeV}$ for opposite-side jets is unchanged
$0 < ~ y_1 < ~ 1$ and $-1 < ~ y_2 < ~ 0$ (and interchange
$y_1 \leftrightarrow y_2$) whereas for same-side jets the rapidity
regions are $0 <  y_1, y_2 < ~ 0.8 (0.4)$ and 
$-0.8(-0.4) < ~ y_1, y_2 <  0 $.
By $E_T > ~ 30 {\rm GeV}$ all regions of rapidity space diminish.
For example at $E_T=40 {\rm GeV}$, the allowed rapidity space is 
$0 < y_1  \stackrel{<}{\sim} 0.5$, $-0.5 \stackrel{<}{\sim} y_2 < 0$
(and $y_1 \leftrightarrow y_2$) and
$0 < y_1, y_2 \stackrel{<}{\sim} 0.2$ and
$-0.2 \stackrel{<}{\sim} y_1, y_2 <  0$.
Finally for $E_T \stackrel{>}{\sim} 45 {\rm GeV}$
there is no allowed rapidity region.

Integrating over the rapidity region in Eq. (\ref{approxn2}) with these
estimates, we find consistency with the
exact numerical results for $d\sigma(E_T)/dE_T$ in 
Figs. \ref{figdpd0}c and \ref{figdpd0}d.
Also the rapid cutoff in the $E_T$-spectra
at $E_T \approx 45 {\rm GeV}$, best seen in Fig. \ref{figdpd0}d,
is consistent with our crude estimates here that the available rapidity
phase space vanishes at this point.

For the N(L)DPE $y_+$-spectra in Fig. \ref{figdpd0}a, 
we again can apply the above
results, except integrating Eq. (\ref{approxn2}) over $E_T$ and $y_-$.
The basic shape of the $y_+$-spectra can be understood by assuming
dominance of the low $E_T$ regime $E_T \sim 15 {\rm GeV}$ and examining
the behavior of the $y_-$ phase space as a function of $y_+$.  For
the latter
we find at $y_+=0$ the $y_-$ range is $\Delta y_-(y_+=0) =4$ and it
vanishes linearly to zero as $y_+ \rightarrow 1$. 
In fact relative to linear decrease as a function of $y_+$,
the $y_+$-spectra in Figs. \ref{figdpd0}a and \ref{figdpd0}b 
is more enhanced near 
the middle $y_+=0$ relative to the periphery $y_+=1$.  This
is an effect of higher $E_T$-dijets. Recall from above that as $E_T$
increases, the rapidity region first to diminish is for
same-side dijets, or equivalently the large $y_+$ region.
For example at $E_T=20 {\rm GeV}$ the region
$1 < |y_+| < ~ 0.8$ no longer has dijets.  In general higher $E_T$ dijets
enhance the region near $y_+=0$ relative to $y_+=1$.

To estimate the N(L)DPE total cross from the approximate expression
Eq. (\ref{approxn2}), the 
$E_T$ integral is performed with the rapidity phase space region
evaluated at $E_T=15 {\rm GeV}$.  The latter yields 
a rapidity area $\sim 4$, which implies for no Sudakov suppression
the estimate
$\sigma \approx 1.7 {\mu {\rm b}}$.
Evaluating the Sudakov suppression factor at $E_T=15 {\rm GeV}$
leads to suppression factors to the total cross section
of $\sim 0.67$ and $\sim 0.11$ for $\mu^2=1 {\rm GeV^2}$ 
and $\mu^2 =0.3 {\rm GeV^2}$ respectively. 
These estimates are within a factor 2 of the exact numerical results in
Table 1.

Turning to the F(IS)DPE process, the steeper $E_T$-spectra relative to both
the N(L)DPE and inclusive processes arises for reasons similar to the
CDF case discussed earlier.  In particular, the high $E_T$ region 
in general requires higher parton momentum fractions, which
are suppressed by the pomeron parton distribution functions.
For $E_T \stackrel{>}{\sim} 22 {\rm GeV}$  both incoming partons 
to the hard process carry "large"
momentum fractions, $x_h/x_{{\pom}/h} > ~ 0.5$.
Thus above this $E_T$, one should expect the $E_T$ spectra
to diminish as evident in Figs. \ref{figdpd0}c and \ref{figdpd0}d.
For example relative to the inclusive $E_T$-spectra,
the F(IS)DPE $E_T$-spectra decreases for the A,B,C,D, SG models
at $E_T=22 {\rm GeV}$ by the additional factors
$6,9,6,9,3$ respectively and at
$E_T=30 {\rm GeV}$ by the additional factors $100,250,100,250,27$
respectively.  This faster decline for the F(IS)DPE
$E_T$ spectra arises because the edge (i.e. $\beta \sim 1$)
of the pomeron parton distribution
functions is being reached. Consistently,
the slowest decline of the $E_T$-spectra amongst the
five models is by the SG model, whose parton distribution
function decreases the slowest
at $\beta \rightarrow 1$.

For the $y_+$-spectra in 
Figs. \ref{figdpd0}a and \ref{figdpd0}b, their shape can be understood
through rapidity phase space consideration and the behavior of the
pomeron parton distribution functions.  Generally larger $|y_+|$ requires
larger $x_p$ and $x_{\bar p}$. Similar to the N(L)DPE case,
at $E_T=20  {\rm GeV}$, for example, the $x_{{\pom}/p,{\bar p}}$
and dijet rapidity cuts prohibit dijets for $|y_+| \stackrel{>}{\sim} 0.8$.
In addition to simple phase space restrictions, for the F(IS)DPE process
larger parton momentum fractions $x_p$ and $x_{\bar p}$ thus
larger $|y_+|$ are further suppressed due to the pomeron parton distribution
functions.  This additional source of suppression at large $|y_+|$ explains
in Figs. \ref{figdpd0}a and \ref{figdpd0}b.
why the F(IS)DPE $y_+$-spectra are a little narrower than the N(L)DPE
ones.
     
Turning to the magnitude of the cross sections,
from Table 1 the relative sizes amongst
the five parton distribution functions are about the same as in the CDF
case discussed earlier, $1(A):50(B):1(C):200(D):50(SG)$.
These ratios can be cross checked with the general behavior of the 
F(IS)DPE cross section formula, similar to our earlier treatment for
the CDF case.  Basically one finds that
these ratios are consistent
with the ratios of the pomeron parton distribution functions in
the typical $\beta$ regime ($\beta \sim 0.1$).

Finally, it is interesting to compare 
cross section magnitudes between D\O\-1800 and CDF.
If the DPE dijet process is N(L)DPE dominated,
the CDF and D\O\-1800 cases differ primarily by phase space area  and the
$1/E_T^2$ scaling factor.  Despite  the larger CDF central rapidity
region, once their $x_{{\pom}/p,{\bar p}}$ cuts are considered, the dijet phase
space area between CDF and D\O\-1800 is approximately the same.
As such the predominant difference between the CDF and D\O\-1800 cross
sections is due to the $1/E_T^2$ factor, which if evaluated at
their respective minima implies the N(L)DPE cross sections
of D\O\-1800 should be a factor $\sim 4$ smaller than those of CDF. This is
consistent with Table 1.

On the other hand, for a F(IS)DPE dominated dijet process,
in addition to the above two factors, an additional difference arises
from the pomeron parton distribution functions.  They make a significant
difference due to their rapid growth at small-$\beta$.
Very roughly, since for CDF $E_T^{\rm min}$ is a factor 2 smaller
and the average $x_{\pom}$ is a factor 2 larger compared to D\O\-1800,
the typical $\beta$ at which the pomeron parton distribution
functions are evaluated  in the CDF case should be a factor 4 smaller
relative to D\O\-1800.  By knowing the typical $\beta$ range for the
two cases, we can estimate the behavior of the parton distribution
functions in that range.  We expect that within
the kinematically accessible range of $\beta$, the typical range that
dominates the cross section will be near its minimum limit since that is
where the parton distribution functions will be the largest.
The smallest $\beta$ possible based on Eq. (\ref{xpmin}) is
$0.015$ and $0.11$ for CDF and D\O\-1800 respectively.  However
at this limit, although one parton distribution function will be large,
this effect is compensated by the other parton distribution function,
which must be evaluated at $\beta=1$ where it vanishes.  Thus
$\beta$ regions away from this limit also will contribute significantly.
As an estimate of an upper bound to the typical $\beta$ range,
we estimate at $E_T^{\rm min}$ for $x_p=x_{\bar p}$  and when $x_{\pom}$ is at
an average value, which we will take as $\sim 0.05$ and $\sim 0.025$ for
CDF and D\O\-1800 respectively.  Then we obtain $x_h = 0.0077$ and $0.017$
so that $\beta =0.15$ and $0.70$ for CDF and D\O\-1800
respectively.  So in summary the typical $\beta$ ranges
are $0.15-0.015$ for CDF and $0.70 - 0.11$ for D\O\-1800.
These estimates confirm that the typical $\beta$ are a factor
$4 - 7$ smaller for the CDF case versus the D\O\-1800 case. 
Moreover the size of $\beta$ for both cuts is
$O(10^{-1})$.  In this $\beta$ range, the growth of 
all five ACTW pomeron parton
distribution functions  is $\sim 1/\beta^{1.2-1.3}$. Thus the typical size
of a pomeron parton distribution function in the CDF case will be
a factor $\sim (4 - 7)^{1.2-1.3}$ bigger than the D\O\-1800 case.
Accounting for this factor for each of the two parton distribution
functions and the factor 4 from $E_T$ scaling, we expect the cross section
for any given pomeron parton distribution function model to be a factor
$\sim 10^2$ larger for CDF relative to D\O\-1800. This essentially is what is
found in Table 1.

Note that the size of the typical $\beta$ found above is interesting in
its own right. It implies the range of $\beta$ probed in both
the CDF and D\O\-1800 cases is not tiny. Recall that very tiny $\beta$ is the
natural regime for gluon dominance. Thus, for the CDF and D\O\-1800 cuts,
one should not assume that gluon dominance is necessary and
we have shown earlier by explicit examples that there are parton distribution
functions models (A and C in particular)
in which that assumption is wrong.

\begin{figure}
   \begin{center}
      \leavevmode
      \epsfxsize=1.0\hsize
      \epsfbox{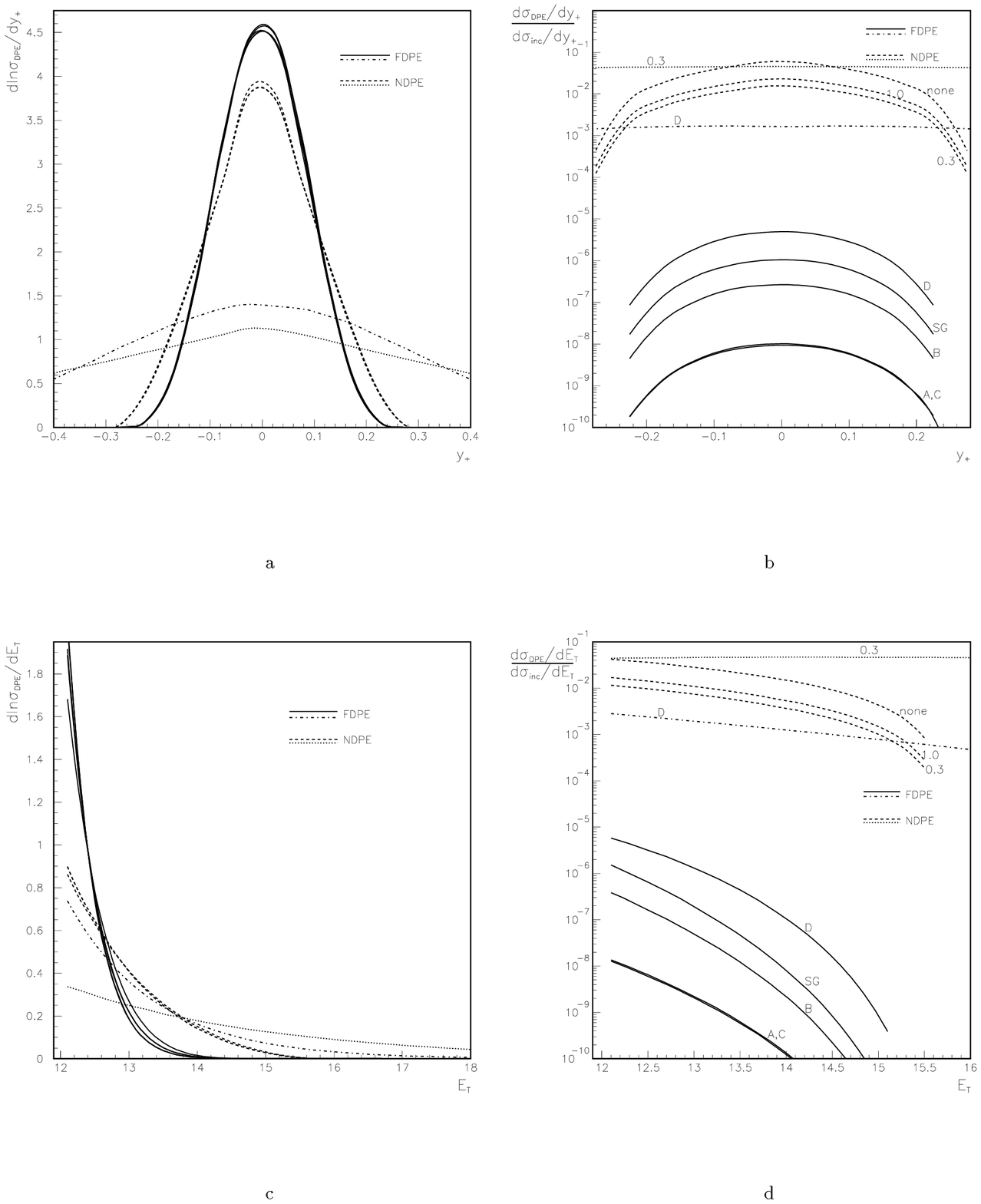}
   \end{center}
   \caption{
       Double pomeron exchange (DPE) dijet production - D\O\-630 cuts. 
(a) Mean
rapidity spectra $y_+ \equiv (y_1+y_2)/2$.
(b) Ratio of mean rapidity spectra between the DPE and corresponding
inclusive dijet processes.
(c) $E_T$-
Spectra. (d) Ratio of $E_T$-Spectra between
the DPE and corresponding
inclusive dijet processes.}
   \label{figdpd0630}
\end{figure}

For D\O\-630, the qualitative features of the spectra basically are the same
as for D\O\-1800.  In Fig. \ref{figdpd0630}a, the $y_+$-spectra for the N(L)DPE
process is a little broader compared to the F(IS)DPE process.
For both the F(IS)DPE and N(L)DPE
processes, the $y_+$-spectra is contained in a much smaller
region, $|y_+| \stackrel{<}{\sim} 0.3$, compared to D\O\-1800.
However, similar to D\O\-1800,
the $E_T$-spectra for the N(L)DPE case falls much slower than for
the F(IS)DPE case.  From Fig. \ref{figdpd0630}d, the relative
decline of N(L)DPE $E_T$-spectra to the inclusive case is
faster than in the D\O\-1800 case.  More noticeably, in comparison 
to D\O\-1800,
the F(IS)DPE $E_T$-spectra falls much faster relative to the corresponding
inclusive $E_T$-spectra. In particular, the ratio of
the F(IS)DPE to the inclusive $E_T$-spectra
for D\O\-630 falls by two orders of magnitude within an increase of $E_T$
by $2 {\rm GeV}$ whereas for the D\O\-1800 case the same decrease
requires an increase of $E_T$ by $15 {\rm GeV}$.  

To understand the features of the N(L)DPE case from 
Eq. (\ref{approxn2}), first note
that at $E_T=12 (15){\rm GeV}$ the allowed rapidity region for
opposite-side dijets is
$0 < y_1 < ~ 0.75 (0.40)$, $-0.75 (-0.40) < ~ y_2 < 0$
(and $y_1 \leftrightarrow y_2$)
and same-side dijets is
$0 < y_1, y_2 < ~ 0.25 (0.10)$, $-0.25 (-0.10) < ~ y_1, y_2 < 0$.
By $E_T \geq 18 {\rm GeV}$ there is no accessible jet rapidity region. 
As such the explicit jet rapidity cuts of $|y_i|< 1$ are irrelevant
since the upper bounds $x_{{\pom}/p,{\bar p}} < 0.05$ already prohibit 
sufficient energy deposition to produce dijets at the higher
$|y_i|$ region at even the lowest permissible $E_T$.  
These crude estimates along with Eq. (\ref{approxn2}) show consistency with
the exact numerical results in Fig. \ref{figdpd0630} and Table 1.  
Finally since the jet rapidity region is rapidly shrinking even at the
lowest $E_T$, this explains the more rapid decline of
of the $E_T$-spectra between the N(L)DPE and inclusive processes
for this case compared to D\O\-1800.  This effect can be reversed by
increasing the upper limit on $x_{{\pom}/p,{\bar p}}$.
For example, the dotted curves in Figs. \ref{figdpd0630}a-d
present the various N(L)DPE spectra for Sudakov suppression
with $\mu^2=0.3$ and when 
$x_{{\pom}/p}, x_{{\pom}/{\bar p}} \leq 0.1$

For the F(IS)DPE process, the explanation for the two spectra essentially is
the same as in the D\O\-1800 case.  The important quantitative difference is
the typical parton momentum fractions are much bigger here than for
D\O\-1800.
This explains the sharper fall of both the $y_+$ and $E_T$ -spectra.
For example, to produce $12 {\rm GeV}$ dijets with the minimal energy
deposition, so at $y_-=0$, requires for symmetric parton momentum fractions
$x_p=x_{\bar p}=0.038$.  Thus almost all the D\O\-630 
events are in the ``large''
$\beta$ regime, $\beta_h \equiv x_h/x_{{\pom}/h} \stackrel{>}{\sim} 0.5$.
Recall in this region the pomeron parton distribution functions are
rapidly diminishing. As such, the sharp decline in the
$E_T$-spectra for D\O\-630 relative to the inclusive case primarily is
because the periphery of the parton distribution functions
at $\beta \sim 1$ are being probed.
This sharp decline can be reduced if the limits
on $x_{{\pom}/p}$ and $x_{{\pom}/{\bar p}}$ are increased.
For example, the dashed-dotted curves in Figs. \ref{figdpd0630}a-d
present the various F(IS)DPE spectra for the ACTW model D
when $x_{{\pom}/p}, x_{{\pom}/{\bar p}} \leq 0.1$

\subsubsection{LHC}

\begin{figure}
   \begin{center}
      \leavevmode
      \epsfxsize=1.0\hsize
      \epsfbox{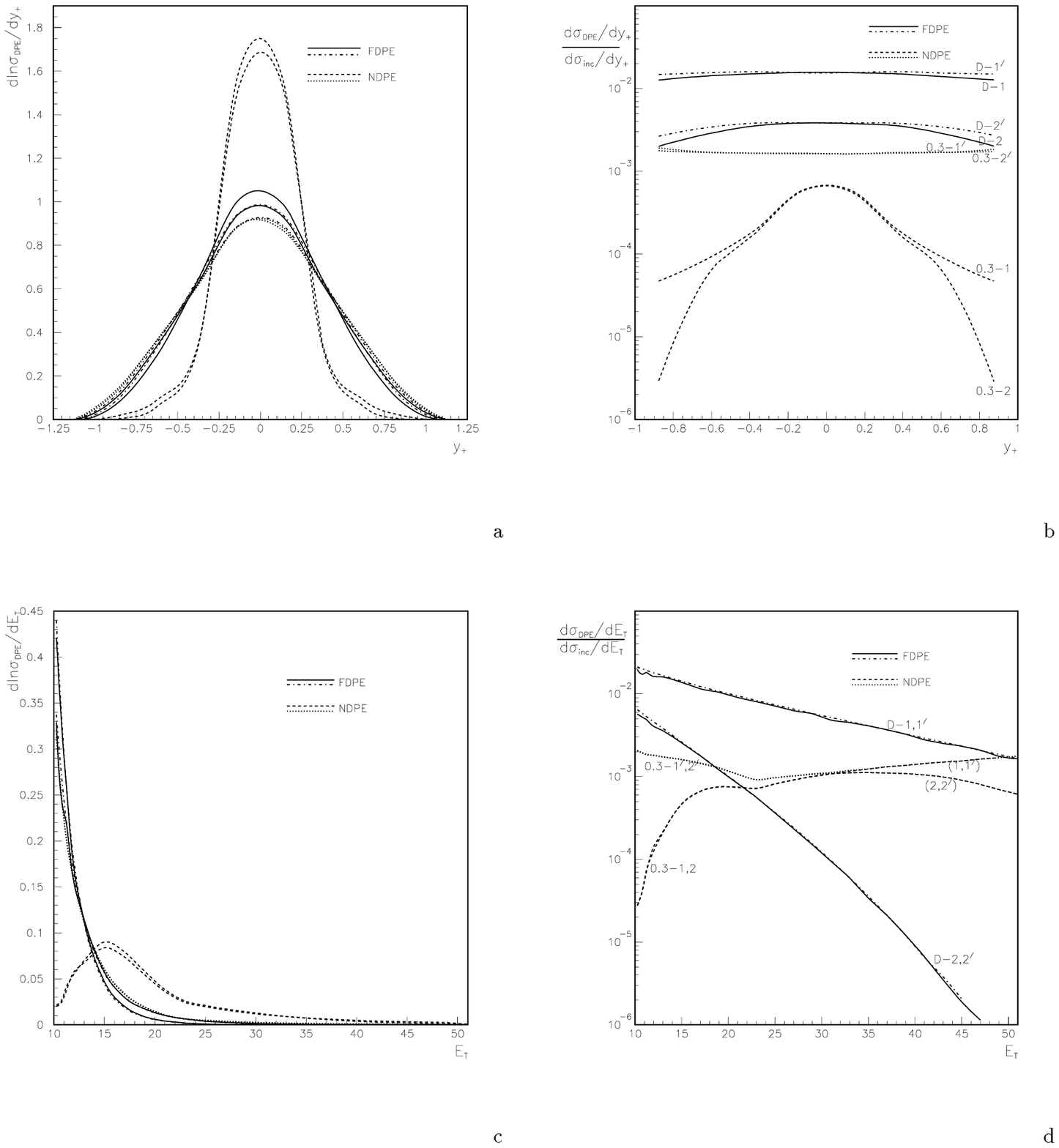}
   \end{center}
   \caption{
       Double pomeron exchange (DPE) dijet production - LHC cuts. (a) Mean
rapidity spectra $y_+ \equiv (y_1+y_2)/2$.
(b) Ratio of mean rapidity spectra between the DPE and corresponding
inclusive dijet processes.
(c) $E_T$-
Spectra. (d) Ratio of $E_T$-Spectra between
the DPE and corresponding
inclusive dijet processes.
   }
   \label{figdplhc}
\end{figure}

The LHC results in Fig. \ref{figdplhc} 
have the following interesting features. In
Figs. \ref{figdplhc}a and \ref{figdplhc}b, 
the $y_+$-spectra for the N(L)DPE LHC-1 and LHC-2
cases (dashed curves) are much 
narrower than for all the other cases.  Relative to the
inclusive $y_+$-spectra in Fig. \ref{figdplhc}b, all the F(IS)DPE cases and
the N(L)DPE LHC-1' and LHC-2' cases (dotted curves) 
are flat, whereas the N(L)DPE LHC-1
and LHC-2 cases drop-off as $|y_+|$ increases with the latter falling
fastest.  For the $E_T$-spectra
in Fig. \ref{figdplhc}c, the most intersting feature is for the
N(L)DPE LHC-1 and LHC-2 cases, for which the $E_T$-spectra actually first
rises with
increasing $E_T$ until $E_T \approx 15 {\rm GeV}$ and thereafter falls.   
In contrast for N(L)DPE LHC-1' and LHC-2', the $E_T$-spectra are of a
more standard behavior.  For the F(IS)DPE $E_T$-spectra, 
the LHC-2 (solid curves) and LHC-2' (dashed-dotted curves)
cases fall much faster with increasing $E_T$ than 
the LHC-1 (solid curves) and LHC-1' (dashed-dotted curves)
cases.  In fact, in Fig. \ref{figdplhc}d
the F(IS)DPE LHC-1 and LHC-1' $E_T$ spectra are almost as
flat as the N(L)DPE LHC-1' and LHC-2' cases.

For the N(L)DPE case, the primary difference between the primed and unprimed
spectra arise due to the different lower bounds
on $x_{{\pom}/p,{\bar p}}$ of $0.00006$ and $0.002$ 
respectively.  Due to the
high $\sqrt{s}$ relative to the CDF and D\O\ cases studied earlier, 
much smaller parton momentum fractions are necessary 
to produce kinematically identical dijets.  In fact the larger lower cut-off
in $x_{{\pom}/p,{\bar p}}$ for the LHC-1 and LHC-2 
cases already is too large to
produce substantial numbers of dijets in the range
$10 {\rm GeV} < E_T \stackrel{<}{\sim} 15 {\rm GeV}$. 
For example, the only rapidity region
where $E_T \approx 10 {\rm GeV}$ dijets can appear is for
opposite-side jets with $0.6 < ~ y_1 < 1$,
$-1.0 < y_2 \stackrel{<}{\sim} -0.6$ (and $y_1 \leftrightarrow y_2$).
Once $E_T$ rises to $15 {\rm GeV}$, the same-side jet region becomes
accessible starting with the region $y_1 \approx y_2 \approx 0$
and moving outward to higher rapidity with increasing $E_T$.
This behavior also explains the narrower $y_+$-spectra for the
N(L)DPE LHC-1 and LHC-2 cases in 
Figs. \ref{figdplhc}a and \ref{figdplhc}b.
In the dominant $E_T$ range 
$10 {\rm GeV} < E_T \stackrel{<}{\sim} 15 {\rm GeV}$,
dijets predominately emerge within $\sim 0.2$ $y_+$-rapidity units
about $y_+=0$.  The shoulder of the $y_+$-spectra 
in Fig. \ref{figdplhc}a at $y_+ \sim 0.4$
corresponds to $E_T \sim 20 {\rm GeV}$.

In contrast, for the LHC-1' and LHC-2' cuts, the lower limit on
$x_{{\pom}/p,{\bar p}}$ imposes no constraints on jet rapidity. In this case,
starting at $E_T=10 {\rm GeV}$, the complete rapidity region
$-1 < y_1, y_2 < 1$ is accessible.  The explanation for the
$y_+$ and $E_T$ spectra in this case follows similar reasoning to the
D\O\-1800 cases discussed earlier.

To compare the magnitude of the N(L)DPE cross sections in Table 1 with
Eq. (\ref{approxn2}), the dijet phase space must be estimated.  For
the N(L)DPE LHC-1' and LHC-2' cases at $E_T=E_T^{\rm min}=10 {\rm GeV}$
the entire explicit rapidity region $-1 < y_1,y_2 < 1$ is accessible
with no additional constraints from
the $x_{\pom}$ cuts.  This obtains the estimate
with no Sudakov suppression
of $\sigma \approx 4 {\mu {\rm b}}$.
For the N(L)DPE LHC-1 and LHC-2 cases, because of the the 
large lower limit
on $x_{\pom}$, at $E_T=10 {\rm GeV}$ dijets only appear
in a small region of opposite-side dijet rapidity space
with $\Delta y_+ \Delta y_- \approx 0.3$.
However at $E_T \approx 15 {\rm GeV}$, the entire opposite-side
jet rapidity region is accessible $0 < y_1 < -1$,
$-1 < y_2 < 0$ (and $y_1 \leftrightarrow y_2$)
but as yet only a negligible region of same-side jet
rapidity space.  Thus the total rapidity area
is $\Delta y_+ \Delta y_- \approx 2$.  Observe that the gain in phase
space area is a factor 3 greater than the suppression factor of $9/4$
from the $1/E_T^2$ behavior of $\sigma$.  This is consistent with
the rise in the $E_T$-spectra in Fig. \ref{figdplhc}a.  Applying the above
estimates to Eq. (\ref{approxn2}) implies for no
Sudakov suppression $\sigma \approx 1 {\mu {\rm b}}$.
The Sudakov suppression factor evaluated at $E_T=10 {\rm GeV}$
implies the cross section in both cases should decrease
by factors $0.4$ for $\mu^2 =1 {\rm GeV}^2$ 
and $0.2$ for $\mu^2 =0.3 {\rm GeV}^2$.
These crude estimates are consistent with the exact numerical results in
Table 1.

For the $y_+$ F(IS)DPE spectra in Fig. \ref{figdplhc}b, 
they are very similar to their
inclusive counterparts. This contrasts the CDF and D\O\ cases in
Figs. \ref{figdpcdf}b, \ref{figdpd0}b, and 
\ref{figdpd0630}b.  The reason is the higher $\sqrt{s}$,
which for fixed $E_T$ and $y_+$, requires smaller parton momentum fractions
$x_{p,{\bar p}}$.  
Assuming the shape of the $y_+$-spectra is dominated by the
lowest $E_T \approx 10 {\rm GeV}$ region, for the full
range of $|y_+| < 1$, $\beta_i \equiv x_h/x_{{\pom}/h}$
typically never is "large"
$\beta_i < 0.5$. Thus for the full
$y_+$ range the pomeron parton distribution functions typically are not
probed near $\beta \rightarrow 1$ where they vanish. Instead,
they typically are probed at intermediate $\beta$ regions, similar to the
situation in the corresponding inclusive case.  This is the basic
reason for the similarity in the $y_+$-spectra between the F(IS)DPE
and inclusive processes.

For the F(IS)DPE $E_T$-spectra from Fig. \ref{figdplhc}d the flatness of
the LHC-1 and LHC-1' cases again arises  from the small
typical  $\beta$ values at which the pomeron parton distribution functions
are evaluated.  The sharper decrease of the F(IS)DPE LHC-2 and LHC-2' cases
arises due to the lower upper bounds on $x_{{\pom}/p,{\bar p}}$
of $0.01$, versus $0.03$, for the LHC-2 and LHC-2' cases.  The main
consequence of these different upper bounds appears in the parton
distribution functions and not from phase space.  This is evident since 
only the latter effect is relevant for the N(L)DPE cases, and for them 
the $E_T$-spectra at large $E_T$ has
no pronounced difference in all 4 cases.
On the other hand, for the pomeron parton distribution functions,
the two different upper bounds on $x_{{\pom}/p,{\bar p}}$ 
imply that for fixed
$E_T$, they are probed in the LHC-2 and LHC-2' cases at typical
$\beta$ values that roughly are 3 times larger relative to
the LHC-1 and LHC-1' cases.  This factor 3 difference has nonnegligible
effect.  Noting that the order of magnitude of the typical $x_h$
is $> O(10^{-3})$ or equivalently
$\beta_h > 0.06$, with this lower bound increasing with $E_T$,
the factor 3 means the difference between $\beta$ close to
$0.1$ where the parton distribution functions are sizable
and $\beta$ closer to $1$ where they vanish rapidly.

\subsection{Comparison with Experiment}
\label{subsect4C}

Both CDF \cite{cdfdd1,cdfdd2,cdfdd3,cdfdd4,cdfnew} and 
D\O\ \cite{d0dd1,d0dd2,d0dd3,d0dd4}
have reported preliminary results on the double
diffractive dijet process.  In Sect. \ref{sect3}
limitations of both
experiments were discussed which prohibit explicit measurement of the
DPE dijet process in the present runs.  Nevertheless,
one likely possibility considered by both experiments is that the majority of
the double diffractive dijet events were DPE.  Under this assumption,
some of the reported features from these experiments will be interpreted
below in terms of the models examined in this paper.

D\O\ reports its $E_T$-spectra for the DPE dijet process to be similar to
the inclusive $E_T$-spectra with comparable 
cuts \cite{d0dd1,d0dd2,d0dd3,d0dd4}.
The earlier CDF preliminary reports \cite{cdfdd1,cdfdd2,cdfdd3,cdfdd4} 
also found this,
although their most recent report \cite{cdfnew} finds
the $E_T$-spectra falls much faster than for the comparable
inclusive $E_T$-spectra.  Based on Figs. \ref{figdpcdf}d, 
\ref{figdpd0}d,\ref{figdpd0630}d,\ref{figdplhc}d, the D\O\ and
earlier CDF $E_T$-spectra lean towards that for the N(L)DPE process.
Amongst the F(IS)DPE processes, the $E_T$-spectra for 
the best fit ACTW model, D, as well as model B differ significantly
from their inclusive counterpart.  However they could be consistent
with the most recent CDF report \cite{cdfnew}.
The F(IS)DPE model with the most similar $E_T$-spectra to the
inclusive process, is ACTW SG.  Recall in this model
the gluon density peaks near $\beta =1 $.

In the D\O\ case, as noted earlier, by increasing the upper bounds on 
$x_{{\pom}/p}$ and $x_{{\pom}/{\bar p}}$, the $E_T$ spectra could be 
more flattened in Figs \ref{figdpd0}d and \ref{figdpd0630}d.
In Figs. \ref{figdpd0} and \ref{figdpd0630}, the various spectra with
upper bounds $x_{{\pom}/p}, x_{{\pom}/{\bar p}} \leq 0.1$
are shown as the dotted curves for the N(L)DPE case with Sudakov suppression
at $\mu^2=0.3$ and the dashed-dotted curves for the F(IS)DPE ACTW D model.
Observe the dashed-dotted 
curves in Figs. \ref{figdpd0}d and \ref{figdpd0630}d
for the ACTW D model are significantly flatter than any of the 
F(IS)DPE cases with upper bounds
$x_{{\pom}/{p,{\bar p}}} < 0.05$ (solid curves)
(for comparison, the dotted curves in these figures are for the
N(L)DPE model with $\mu^2=0.3$ and 
$x_{{\pom}/{p,{\bar p}}} < 0.1$). 
Since D\O\ does not explicitly measure $x_{{\pom}/p,{\bar p}}$, 
one explanation for
their $E_T$-spectra is the F(IS)DPE model with the higher upper limits
on $x_{{\pom}/{p, {\bar p}}}$, say between $0.05 - 0.1$.
In such a case, recall that the
interpretation of pomeron dominated exchange enters into question.
For D\O\-1800 larger upper limits
on $x_{{\pom}/p,{\bar p}}$ are inconsistent with the largest
DPE dijet $E_T$ that is found, $\sim 52$ GeV. This maximum is
consistent with an upper bound of only $x_{{\pom}/{p, {\bar p}}} < 0.057$.
On the other hand, for D\O\-630, note from \cite{d0dd1,d0dd4}
that dijets are produced up to $E_T \sim 25$ GeV. If background 
and resolution effects or any other experimental complication can be
ruled out, this suggests for D\O\-630
$x_{{\pom}/{p, {\bar p}}} \stackrel{<}{\sim} 0.08$.

For the $y_+$-spectra, the CDF results unambiguously contradict
interpretation as N(L)DPE dominated.  The CDF $y_+$-spectra is very
broad, ranging as 
$-2 \stackrel{<}{\sim} y_+ \stackrel{<}{\sim} 1.5$. 
From Fig. \ref{figdpcdf}a, this
is similar to the F(IS)DPE cases and is completely at odds with the
N(L)DPE cases.
Recall for the CDF cuts, the narrower $y_+$-spectra for the N(L)DPE
model is intrinsic to its lossless nature.  Thus the experimental
$y_+$-spectra is strong indication that the F(IS)DPE process dominates
in the CDF case.  This interpretation also is consistent for the 
$E_T$-spectra in 
the most recent CDF report \cite{cdfnew}.  However, for the
$E_T$-spectra from the earlier CDF reports 
\cite{cdfdd1,cdfdd2,cdfdd3,cdfdd4},
ACTW SG is the most consistent model.
Finally, one can not exclude the possibility that the N(L)DPE process
gives a nondominant but measurable effect.  This would help flatten
any of the F(IS)DPE models in Fig. \ref{figdpcdf}d. 
However an admixture of the
N(L)DPE process also will imply greater enhancement of the
$y_+$-spectra in the region 
$-1 \stackrel{<}{\sim} y_+ \stackrel{<}{\sim} 0$.

Finally, total cross sections can be compared.  CDF has
preliminarily reported a DPE dijet cross section
of $\sigma_{DPE}^{CDF} \approx 13.6$ nb \cite{cdfdd1}. D\O\ has given no
preliminary cross section, although for D\O\-1800 they have
estimated $\sigma_{DPE}^{D\O\ } \sim 10$ pb \cite{d0dd2}.
In this case, the CDF total cross section is about 1000 times larger
than D\O\-1800. The model that most closely obtains this factor
difference is F(IS)DPE ACTW D, which predicts the ratio 
to be still a factor 5 smaller.  As another case, for ACTW SG,
it predicts the CDF total cross section should be only a factor 
$\sim 50$ larger than D\O\-1800.  Furthermore, the predicted ratio
between the CDF and D\O\-1800 total cross sections decreases if
the limits on $x_{{\pom}/{p,{\bar p}}}$ for D\O\-1800 are increased.
Thus there appears to be some discrepancy between the experimental
cross sections and those predicted by all the F(IS)DPE models.
Also, the N(L)DPE models do very poorly in predicting the observed
ratio.  For the largest ratio predicted by these models,
the CDF total cross section would be only a factor 7
larger than D\O\-1800.  However for the N(L)DPE case,
modifications of the proton form factor and two-gluon pomeron model,
which are discussed in the Conclusion,
may change these predictions by an order of magnitude but not more.

\section{Conclusion}
\label{sect5}

This paper has examined the factorized Ingelman-Schlein
and nonfactorized lossless DPE processes of dijet production
and computed predictions  for the cuts of
CDF, D\O\, and representative cuts of LHC.
Two qualitative features emerge from our calculations
which are model independent. They are reflections of the lossless
kinematics of jet production in the
N(L)DPE process, which requires that all
the momentum carried by the pomerons will go into
the hard event. The first of these qualitative differences
between the N(L)DPE and F(IS)DPE processes emerges in
the CDF $y_+$-distributions in Figs. \ref{figdpcdf}a and \ref{figdpcdf}b. 
For this case, it
is evident that for the N(L)DPE process,
the distribution is considerably localized
to within one unit of $y_+$ rapidity whereas for the F(IS)DPE
process, its distribution is considerably broader.
The difference in the shapes of the distributions for
these two processes is due to the intrinsic differences
in the hard kinematics for the two processes and thus
is expected to survive any nonperturbative modifications
of the basic models. 
A second qualitative difference emerges
from the lower bound dependence of $x_{\pom}$ for the 
N(L)DPE process.  In our calculations, this difference
was explicitly seen only amongst the set of LHC
cuts, although the basic feature is general.  Because of the 
lossless kinematics, if the lower bound on
$x_{\pom}$ is sufficiently large, low-$E_T$ jets will be prohibited
from productions.  An example of this is seen
in the LHC $E_T$-spectra, 
Figs. \ref{figdplhc}c and \ref{figdplhc}d, where for
the $1'$ $2'$ cuts ($x_\pom > 0.00006$) no truncation
of low-$E_T$ jets is seen, whereas for the 1,2 cuts,
where the lower bound on $x_{\pom}$ is larger, $x_{\pom} > 0.002$,
a truncation of low-$E_T$ jets occurs.

Two additional qualitative features were found from
our calculations which have some degree of model dependence,
but for which we expect the general trends to sustain.
The first of these, which appears for all the cuts,
is that the $E_T$ spectra are flatter (or ``harder'') for comparable
N(L)DPE versus  F(IS)DPE processes.
This difference occurs due to the additional
$\beta$-dependence in the F(IS)DPE process that arises from the
pomeron parton distribution functions.  At fixed
$\sqrt{s}$, larger $E_T$ implies larger $\beta$, and the
parton distribution functions fall, generally quickly,
with increasing $\beta$.
There are modifications \cite{Martin:1997kv} to the basic N(L)DPE model, 
which will be discussed below, that can introduce a similar type
of $\beta$-dependence.  As such, this difference between
the N(L)DPE and F(IS)DPE processes has some model dependence.
The second qualitative difference is in regards to the
total cross sections, in particular the ratio
of the CDF to D\O\-1800 total cross sections. 
The N(L)DPE process generally has a much smaller ratio for
$\sigma_{CDF}/\sigma_{D\O\-1800}$ $\sim 4$ compared to
the F(IS)DPE process for which this ratio
is typically $\sim 100$.  However, any modeling
that relies on the hard kinematics can alter
this result, such as the $\beta$-dependent modeling
in \cite{Martin:1997kv}.
 
In regards to our comparisons with the preliminary experimental data,
we feel no final conclusions are possible, but the trends in
the data show slight preference for 
domination by the F(IS)DPE process.
From comparison of our models with either the recent \cite{cdfnew} or 
earlier \cite{cdfdd1,cdfdd2,cdfdd3,cdfdd4} CDF 
preliminary data, the mean rapidity ($y_+$)
spectra provide the most suggestive evidence that this
data is dominated by the F(IS)DPE process.  On the other
hand, the D\O\ data shows no greater preference for either
of the two mechanisms.  Furthermore, none of the theoretical
models examined here are completely consistent with all the data.
In particular, attempting to conclude that the data
is dominated by the F(IS)DPE process is
inconsistent with the harder $E_T$-spectra found in the
D\O\ data and the earlier CDF data, since 
such spectra do not readily agree with
the best fit ACTW F(IS)DPE model.
Also, and perhaps most interesting, the experimental ratio of the CDF to
D\O\-1800 total cross sections is a factor $5 - 10$ larger than
predicted by any of the models examined here. 
In this respect, it is worth noting that from Table 1, ratios between
the D\O\-630 and D\O\-1800 cross sections also
can be obtained, and once D\O\ complete their analysis of
their preliminary data, this may be the next test
between the predicted 
ratios and experiment.
Finally, although it appears there are more indications that the data
is dominated by the F(IS)DPE process,
for the N(L)DPE process, neither data nor our analysis is sufficiently
precise to rule out a subdominant component. 
All these issues are important to resolve as further experimental data
becomes available.
For the time being, the limitations of both the CDF
and D\O\ experiments for this process exclude any final conclusions to
be drawn.

Our study emphasized the hard physics, but made minimal attempt at
modeling most of the nonperturbative soft physics.  For example,
general belief is hard, pure hadron induced, diffractive processes
are subject to a weakly $\sqrt{s}$-dependent
suppression factor \cite{regmod,Bjorken:1993er},
which represents the probability for the rapidity gap(s) not to be
filled by extra exchanges of pomerons and gluons between the particles
in the model which have very different rapidities.
Such absorptive corrections potentially can be treated by the methods
developed in \cite{regmod}.  In \cite{Martin:1997kv}, they used
the approach of \cite{regmod} and \cite{kmrabsorb} to estimate the absorptive
correction factor for the DPE dijet process and at
$\sqrt{s} = 1800$ GeV found it to be $\sim 0.06$.  For our calculations,
this factor is meant to
multiply the results in Table 1.  

For the N(L)DPE model, also not treated here is modifications of the
two gluon exchange model with LLA ladder evolution of gluons.
A model for this was given in \cite{Martin:1997kv}.
Their model amounts to including in the N(L)DPE dijet amplitude
two factors of gluon densities evaluated at
$x_{{\pom}/p}$ and $x_{{\pom}/{\bar p}}$.
This modification has interesting consequences.  For example,
maintaining the same procedure to normalize the DPE dijet
amplitude with respect to the elastic $p {\bar p}$ cross section,
the N(L)DPE cross section with this modification
then will be $1-2$ orders of magnitude
smaller.
This arises because the normalization constant is
fixed for elastic scattering process,  where 
$x_{{\pom}/p,{\bar p}}^{\rm elastic} \sim {\bf l}_{\perp}/\sqrt{s}$ 
is small
since ${\bf l}_{\perp}$, the transverse momentum of the outgoing
hadrons, is $\sim 1$ GeV.  For example at $\sqrt{s}=1800$ GeV,
$x_{{\pom}/p,{\bar p}}^{\rm elastic} \sim 0.0006$.
On the other hand, for the
N(L)DPE process the typical 
$x_{{\pom}/p,{\bar p}}$ are $\sim 0.01$.
These are much larger, which implies a
decrease of the
gluon densities relative to the elastic scattering situation. 
This provides an additional suppression
compared to the same model with no gluon densities.  
In \cite{Martin:1997kv},
the exclusive double diffractive dijet model is the same as
our N(L)DPE model here, except for the inclusion of
gluon densities in their model.
It is due to the effect of these densities 
that their model predicts cross sections
$1-2$ orders of magnitude smaller than ours.

Another effect of the gluon densities is the N(L)DPE $E_T$-spectra
will fall faster with $E_T$. The reasons for this are the same as
explained
in Sect. \ref{sect4} for the F(IS)DPE models.  In short, higher $E_T$
requires larger $x_{{\pom}/p,{\bar p}}$ which in turn implies smaller
gluon densities.  This effect will be less pronounced for the N(L)DPE
models versus the F(IS)DPE models, since the argument of the gluon
densities in the former always remain small,
$\stackrel{<}{\sim} x_{\pom}^{\rm max} \sim 0.1$
whereas in the latter it ranges up to $\sim 1$, where the maximum
diminution of the gluon densities occurs.

In summary, this paper examined two very different 
types of diffractive mechanisms
for DPE dijet production, the factorized Ingelman-Schlein and
nonfactorized lossless mechanisms.  In the spirit of Regge physics,
both mechanisms were termed double pomeron exchange.
Some of the differences between the two processes have been elucidated
here, which will help in interpreting experiment.
The F(IS)DPE model appears best to represent the present experimental
data.  However, inconsistencies still remain that need to be sorted-out
by both theory and experiment before final conclusions can be made.
Although the N(L)DPE process does not appear to dominate the cross
section, a subdominant component of it can not be excluded
with present information.  The cleanliness of the final state in this
process, two outgoing hadrons plus a hard event, suffices justification
to search for it.  For new particle search experiments, this process
could permit the ultimate measurement, although its diminutive cross
section precludes it from being the ideal measurement.  Thus as should be
the case with any good tale, the hard double pomeron exchange story is
filled with uncertainties and conflicts.  Furthermore, the hope vested
in the N(L)DPE process someday may vindicate a 
familiar moral, that the best
things don't come easy.

\section*{Acknowledgments}

I thank the following for helpful discussions:
L. Alvero, J. Collins, M. Strikman, J. Whitmore, R. Hirosky,
T. Taylor-Thomas, M. Albrow, K. Goulianos, P. Melese, and K. Terashi.
I also thank L. Alvero for use of his pomeron 
parton distribution codes and for contributions 
to the earlier developments of this paper.
This work was supported in part by the U.S. Department
of Energy.



\newpage

\textwidth 6.5in
\oddsidemargin -0.5in
\evensidemargin -0.5in

\begin{center}
\begin{tabular}{c|ccccc|ccc}
&  \multicolumn{5}{c}{$\sigma^{dijet}_{F(IS)DPE}$(${\rm \mu b}$)}
& &   {$\sigma^{dijet}_{N(L)DPE}$(${\rm \mu b}$)}   \\
\cline{2-9}
                      & {\rm ACTW} & {\rm ACTW} &
{\rm ACTW} & {\rm ACTW} & {\rm ACTW} & No & Sudakov & Sudakov \\
& A & B & C & D & SG & Sudakov & Suppression & Suppression \\
 {cuts} & & & & & & Suppression & $ \mu^2=1.0$ & $\mu^2=0.3$ \\ 
\hline
CDF   & $0.011$ & $1.0$ & $0.011$ & $4.2$ & $0.24$ & $5.6$ & $2.5$
& $1.7$
 \\
D\O\-630   &$2.6\times10^{-8}$ &$7.5\times10^{-7}$ &
$2.9\times10^{-8}$ &$2.9\times10^{-6}$ &$1.4\times10^{-5}$ 
& $0.19$ & $0.075$ & $0.051$
 \\
D\O\-1800   &$9.5\times10^{-5}$ &$5.2\times10^{-3}$ &
$9.9\times10^{-5}$ &$2.2\times10^{-2}$ &$5.4\times10^{-3}$ 
& $1.5$ & $0.37$ & $0.25$
 \\
LHC-1 & $0.027$ & $2.6$ & $0.025$ & $11$ & $0.42$ & $1.6$ &
$0.37$ & $0.25$  
\\
LHC-1' & $0.028$ & $2.7$ & $0.027$ & $11$ & $0.49$ & $5.1$ & 
$1.8$ & $1.2$ 
\\
LHC-2 & $0.0068$ & $0.59$ & $0.0068$ & $2.4$ & $0.15$ & $1.4$ & 
$0.35$ & $0.23$ 
\\
LHC-2' & $0.0075$ & $0.64$ & $0.0075$ & $2.6$ & $0.19$ & $5.0$ &
$1.7$ & $1.2$ 
\\
\end{tabular}

\vspace{5 mm}
\end{center}

Table 1: Double Pomeron Exchange (DPE) dijet total cross sections
for the nonfactorized (N) and factorized (F) models. 

\newpage
\oddsidemargin 0in
\evensidemargin 0in


\vspace{0.5cm}


\pagestyle{empty}

\newpage

\newpage

\newpage

\newpage

\end{document}